\documentclass[a4paper,11pt]{article}
\usepackage{jheppub} 
\usepackage{orcidlink}

\usepackage{mathtools}
\usepackage{amsmath}
\usepackage{amsfonts}
\usepackage{amssymb}
\usepackage{bbold}
\usepackage{bm}
\usepackage{bbm}
\usepackage{esint}
\usepackage{booktabs}    
\usepackage{orcidlink}

\newcommand{\bD}{{\bf D}}
\newcommand{\br}{{\bf r}}
\newcommand{\bn}{{\bf n}}
\newcommand{\bk}{{\bf k}}
\newcommand{\bK}{{\bf K}}
\newcommand{\bv}{{\bf v}}
\newcommand{\bp}{{\bf p}}
\newcommand{\eln}{\rm DLN}

\newcommand{\tomegaE}{{\tilde{\omega}_{E}}}

\newcommand{\omegaE}{{\omega_{E}}}

\newcommand{\sA}{{\sf A}}

\newcommand{\sH}{{\sf H}}
\newcommand{\sM}{{\sf M}}
\newcommand{\sN}{{\sf N}}
\newcommand{\sP}{{\sf P}}
\newcommand{\sB}{{\sf B}}
\newcommand{\sC}{{\sf C}}
\newcommand{\oG}{{\overline{G}}}
\newcommand{\opsi}{{\overline{\psi}}}

\newcommand{\bpartial}{\boldsymbol{\partial}}
\newcommand{\tV}{\theta_{\rm V}}

\long\def\exclude#1{}

\newcommand{\GF}{G_{\rm F}}

\title{Dispersion relation of the neutrino plasma:\\
Unifying fast, slow, and collisional instabilities\footnote{This post-publication arXiv version corresponds to the published version \href{https://doi.org/10.1007/JHEP01(2026)147}{{\em JHEP} {\bf 01} (2026) 147} combined with a published Erratum 
\href{https://doi.org/10.1007/JHEP11(2026)067}{{\em JHEP} {\bf 11} (2026) 067}, see p.~55--57.}}

\author[a]{Damiano F.\ G.\ Fiorillo \orcidlink{0000-0003-4927-9850}} 
\affiliation[a]{Deutsches Elektronen-Synchrotron DESY,
Platanenallee 6, 15738 Zeuthen, Germany}

\author[b]{and Georg G.\ Raffelt
\orcidlink{0000-0002-0199-9560}}
\affiliation[b]{Max-Planck-Institut f\"ur Physik, Boltzmannstr.~8, 85748 Garching, Germany}

\abstract{In neutrino-dense astrophysical environments, these particles exchange flavor through a coherent weak field, forming a collisionless neutrino plasma with collective flavor dynamics. Instabilities, which grow and affect the environment, may arise from neutrino-neutrino refraction alone (fast limit), vacuum energy splittings caused by masses (slow limit), or neutrino-matter scattering (collisional limit). We present a comprehensive analytical description of the dispersion relation governing these unstable modes. Treating vacuum energy splittings and collision rates as small perturbations, we construct a unified framework for fast, slow, and collisional instabilities. We classify modes into \textit{gapped}, where collective excitations are already present in the fast limit but rendered unstable by slow or collisional effects, and \textit{gapless}, which are purely generated by these effects. For each class, we derive approximate dispersion relations for generic energy and angle distributions, which reveal the order of magnitude of the growth rates and the nature of the instabilities without solving directly the dispersion relation. This approach confirms that slow and collisionally unstable waves generally grow much more slowly than they oscillate. Consequently, the common fast-mode approximation of local evolution within small boxes is unjustified. Even for fast modes, neglecting large-distance propagation of growing waves, as usually done, may be a poor approximation. Our unified framework provides an intuitive understanding of the linear phase of flavor evolution across all regimes and paves the way for a quasi-linear treatment of the instability’s nonlinear development.
}

\begin{document}

\renewcommand{\thefootnote}{$\dagger$}
\maketitle

\renewcommand{\thefootnote}{\arabic{footnote}}

\flushbottom

\section{Introduction}
\label{sec:intro}

In neutrino-dense environments, most notably core-collapse supernovae (SNe) and neutron-star mergers (NSMs), neutrino-neutrino scattering is typically subdominant in transporting energy and lepton number, as well as in driving flavor evolution. Instead, weak interactions with other particles, especially charged-current processes involving nucleons, provide the dominant contribution. Surprisingly, however, neutrino-neutrino refraction \cite{Pantaleone:1992eq} can lead to significant flavor redistribution in phase space, despite the collisionless nature of the neutrino gas \cite{Samuel:1993uw, Samuel:1995ri, Duan:2005cp, Duan:2006an, Tamborra:2020cul, Johns:2025mlm}. This form of collective flavor evolution can occur even without flavor mixing \cite{Sawyer:2004ai, Sawyer:2008zs, Chakraborty:2016lct, Izaguirre:2016gsx, Capozzi:2017gqd}, and has attracted considerable attention in recent years; for a representative list of recent works by different groups see Refs.~\cite{Bhattacharyya:2020jpj, Nagakura:2022kic,  Xiong:2023vcm, Shalgar:2022lvv, Cornelius:2023eop, Froustey:2024sgz, Abbar:2024ynh, Abbar:2024nhz, Richers:2024zit, Kost:2024esc, Fiorillo:2024qbl, Nagakura:2025brr, Liu:2025tnf, Padilla-Gay:2025tko}. It remains a missing ingredient in consistent numerical simulations of stellar collapse and NSMs, yet may have substantial astrophysical consequences \cite{Ehring:2023lcd, Ehring:2023abs,Nagakura:2023mhr, Wu:2017drk, Wang:2025nii, Qiu:2025kgy}. Although the Boltzmann collision equation for neutrino flavor evolution at the mean-field level is well established \cite{Dolgov:1980cq, Rudsky, Sigl:1993ctk, Fiorillo:2024fnl, Fiorillo:2024wej}, solving it directly is currently unfeasible due to the vast range of relevant dynamical scales.

In a long series of papers~\cite{Fiorillo:2024bzm, Fiorillo:2024uki, Fiorillo:2024pns, Fiorillo:2025ank}, we have recently developed a new perspective on this topic, inspired by its analogy to collective vibrations in a \textit{collisionless plasma}, that is, an electron-ion plasma in which electrons, despite not colliding directly, exchange energy through their electric fields \cite{vlasov1968vibrational, Landau:1946jc, schKT, fitzpatrick2022plasma}. Likewise, in a dense yet largely collisionless neutrino gas, the particles can exchange energy -- albeit not very efficiently due to the smallness of the refractive energy shift -- and flavor through collective flavor waves (\textit{flavomons}~\cite{Fiorillo:2025npi}), analogous to the plasma waves (\textit{plasmons}), that were systematically studied decades ago by a completely different community, in a completely different context.

Similar to a collisionless electron plasma, the weak field that enables flavor exchange among neutrinos is typically very feeble and does not cause large effects. However, under certain circumstances, the neutrino energy and angular distributions become unstable, leading to an exponential growth of the weak field and, in turn, to large flavor conversions. Such instabilities generally signal a transition to a state of lower free energy, with the excess energy being shed as a collection of flavor waves or, equivalently, as their quanta, the flavomons \cite{Fiorillo:2025npi}. In other words, the instability can be seen as evidence of neutrinos carrying excess energy that is released as flavomons. An obvious conclusion from this viewpoint is that a neutrino plasma is unstable only when it deviates from thermal and chemical equilibrium, since otherwise it would already minimize its free energy. Collective flavor conversions drive the system toward equilibrium, but strong constraints imposed by their collisionless nature typically prevent complete equilibrium.

This perspective also sheds light on an otherwise mysterious feature: neutrino-matter collisions can produce novel classes of instabilities~\cite{Johns:2021qby, Xiong:2022zqz, Liu:2023pjw, Lin:2022dek, Johns:2022yqy, Padilla-Gay:2022wck, Fiorillo:2023ajs}. This effect has long seemed counterintuitive, since standard neutrino oscillations are hindered by collisions, which damp flavor coherence. However, once flavor conversions are understood as a mechanism by which neutrinos relax to more stable states -- emitting excess energy in the form of flavomons -- it becomes clear that collisions can sometimes trigger instabilities. This is because they remove the constraint of entropy conservation, thereby facilitating the system’s approach to states of lower free energy.

Therefore, two key questions emerge that drive the phenomenology of collective flavor conversions: Which conditions cause a neutrino plasma to become unstable, and what state does the system evolve to? In this work, we focus exclusively on the first question. Recent progress has primarily covered the \textit{fast flavor} case \cite{Chakraborty:2016lct, Izaguirre:2016gsx, Capozzi:2019lso, Yi:2019hrp}, defined by neglecting neutrino masses and collisions entirely. The main simplification is the reduction to a single dimensional parameter: the neutrino-neutrino refractive energy shift. On this basis, we have fully characterized the fast flavor dispersion relation \cite{Fiorillo:2024bzm, Fiorillo:2024uki, Fiorillo:2024dik}, notably deriving approximate expressions for the growth rates across all regimes -- limited only by the assumption of a homogeneous neutrino background. This simplification is justified for modes with wavelengths much shorter than the inhomogeneity scale.

In contrast, no such comprehensive study exists once neutrino masses and collisions are included. While our recently developed theory of \textit{slow flavor} conversions (defined by the inclusion of neutrino masses) was extensive \cite{Fiorillo:2024pns, Fiorillo:2025ank}, it was restricted to a monochromatic energy distribution. Very recently, the nonlinear evolution of this case and its approach to asymptotic configurations was examined numerically in Ref.~\cite{Padilla-Gay:2025tko}, but limited to  vacuum frequencies so large that slow instabilities dominate over fast ones. We had identified this regime in Ref.~\cite{Fiorillo:2024pns} and concluded that it is not likely to represent realistic astrophysical environments. Additionally, neutrino-matter collisions can trigger new forms of instabilities \cite{Johns:2021qby}, which were systematically explored in both the linear \cite{Xiong:2022zqz, Liu:2023pjw, Lin:2022dek, Johns:2022yqy} and nonlinear regimes \cite{Padilla-Gay:2022wck, Fiorillo:2023ajs}, but only under the restrictive assumption of perfect homogeneity and isotropy (one exception is Ref.~\cite{Liu:2023pjw}, which considers a few numerical examples of inhomogeneous modes, but the extent of generality of these results is not discussed). While these treatments parallel the fast \cite{Johns:2019izj, Padilla-Gay:2021haz} and slow \cite{Hannestad:2006nj} flavor pendulum models, which have been fully understood in Refs.~\cite{Fiorillo:2023mze, Fiorillo:2023hlk} and \cite{Raffelt:2011yb, Yuzbashyan:2018gbu}, respectively, such highly symmetric systems are unlikely to yield general insights applicable to more realistic scenarios.

Our primary goal is to develop a comprehensive understanding of flavor instabilities in the fast, slow, and collisional limits. We aim for a unifying perspective: rather than simply extending the theory of slow flavor conversions to general energy distributions, or formulating a separate theory for collisional flavor conversions, we treat neutrino masses and collisions on equal footing. This unified approach is more illuminating because it highlights both the similarities and differences among these regimes. This unification is possible because both the vacuum frequency and collisions act as small perturbations to the fast regime. Accordingly, we can construct a perturbative framework that connects the fast regime with the slow and collisional limits.

Specifically, our strategy is to construct a series of approximations, each valid within a distinct regime, to capture the qualitative behavior of unstable and damped modes without relying on black-box numerical methods.  Rather than presenting numerical solutions across a broad parameter space -- as is common in previous studies -- we focus on identifying the generic features of the dispersion relation that govern growth-rate scalings and reveal the interplay among different terms. We restrict our analysis to distributions without angular crossings. Otherwise, fast instabilities typically dominate and are expected to relax rapidly, driving the distribution toward a form without angular crossing. Fast instabilities have already been fully characterized in Refs.~\cite{Fiorillo:2024bzm, Fiorillo:2024uki}.

We begin our exposition in Sec.~\ref{sec:dispersion_relation} with a recap of the dispersion relation of the neutrino plasma in its most general form, and reduce it to the axisymmetric two-flavor case that forms the basis of our work. In Sec.~\ref{sec:scales}, we discuss the key parameters,  characteristic energy scales, and their orders of magnitude. This understanding is essential, since our key method is a perturbative expansion based on the hierarchy of energy scales, and our main result is to develop consistently the orders of magnitude of the growth rate. In Secs.~\ref{sec:superluminal}--\ref{sec:gapless}, we discuss four different types of modes which we newly classify: superluminal, subluminal, near-luminal (based on their phase velocity, which determines whether they can resonantly interact with neutrinos), and gapless modes, which exist only by virtue of neutrino masses and collisions. In Sec.~\ref{sec:example_distributions}, we show how our general results emerge from the practical numerical solution of the dispersion relation for specific examples of neutrino distributions. In Sec.~\ref{sec:thermodynamics}, we demonstrate how collisional instabilities fundamentally differ from fast and slow instabilities due to their tendency to decrease the free energy. We also prove the H-theorem for collisional instabilities, since the original proof~\cite{Sigl:1993ctk} does not apply here. Finally, in Sec.~\ref{sec:discussion}, we summarize our findings, discuss how generic they are, and outline how they can serve as a starting point for investigations of non-linear flavor evolution in realistic environments.

\section{General form of the dispersion relation}\label{sec:dispersion_relation}

In this section, we review the well-known kinetic equations, their linearization, and the associated dispersion relation. While we do not introduce new material here, we establish the formal framework and notation and provide a classification of unstable modes as fast, slow, and collisional, as well as a new classification of gapped vs.\ gapless.

\subsection{Boltzmann kinetic equation}

The flavor dynamics is expressed in terms of the usual neutrino $3\times 3$ flavor density matrix $\rho(\bp,\br,t)$, which depends on the neutrino momentum $\bp$, position $\br$, and time $t$. The diagonal elements are the familiar occupation numbers $f_\alpha(\bp,\br,t)$ for neutrino flavor $\nu_\alpha$. The off-diagonal elements represent the amount of coherence between flavors $\alpha$ and $\beta$. Throughout, we work in the weak-interaction basis, not the mass basis. To include antineutrinos, we do \textit{not} use the flavor-isospin convention; instead the diagonal elements of $\overline{\rho}(\bp,\br,t)$ are the \textit{positive} occupation numbers $\bar f_\alpha(\bp,\br,t)$. 

In the mean-field limit, the evolution is governed by the well-known quantum-kinetic equation \cite{Dolgov:1980cq, Rudsky, Sigl:1993ctk, Fiorillo:2024fnl, Fiorillo:2024wej}
\begin{equation}\label{eq:QKE}
    (\partial_t+\bv\cdot\bpartial_\br)\rho(\bp,\br,t)
=-i\bigl[\sH(\bp,\br,t),\rho(\bp,\br,t)\bigr]
+{\sC}(\rho,\overline\rho).
\end{equation}
On the left-hand side, the Vlasov operator describes the streaming of ultrarelativistic neutrinos with velocity $\bv=\bp/|\bp|$. Here we neglect the weak force produced by the inhomogeneous flavor composition of the neutrinos, even though in the presence of flavor instabilities such forces are conceptually important since they lead to the non-conservation of the weak interaction energy among neutrinos; this is discussed in Ref.~\cite{Fiorillo:2024fnl}.
The refractive Hamiltonian matrix is
\begin{equation}\label{eq:Hamiltonian}
    \sH(\bp,\br,t)=\pm\frac{\sM^2}{2E}+\sqrt{2}\GF\sN
    +\sqrt{2}\GF\int\frac{d^3\bp'}{(2\pi)^3}\,\bigl[\rho(\bp',\br,t)-\overline\rho(\bp',\br,t)\bigr](1-\bv'\cdot\bv).
\end{equation}
The mass matrix $\sM$ provides the vacuum energy splitting among neutrinos and antineutrinos with different energy $E=|\bp|$. The upper sign provides $\sH$ for neutrinos, the lower sign $\overline\sH$ for antineutrinos. The matrix $\sN$ of net charged fermion densities (i.e., it has $n_{e^-}-n_{e^+}$ etc.\ on the diagonal) provides the refractive potential produced by background matter. This expression assumes the absence of a matter current, and $\GF$ is Fermi’s constant. Most important is neutrino-neutrino refraction embodied by the third term which depends self-consistently on the neutrino and antineutrino density matrices.

Finally, $\sC(\rho,\overline{\rho})$ is the collision term for neutrinos and antineutrinos with matter and with each other. Collisions are often ignored for pure collective-oscillation studies, but of course, they shape the neutrino distributions that will or will not show instabilities and nontrivial flavor evolution. In addition, damping of flavor coherence surprisingly can cause another class of instabilities~\cite{Johns:2021qby} beyond the more traditional fast and slow modes. The most general collision term is quite complicated, involving various types of pair processes as well as neutrino-neutrino collisions. However, the main source of neutrino opacity consists of neutral and charged current interaction with nuclei. 

Therefore, as a main mechanism to damp flavor coherence, we consider $\nu_e$ and $\overline{\nu}_e$ beta processes, while ignoring them for muon-flavored states because of the relative scarcity of muons. In other words, the collision term reduces to a source and sink term for $\nu_e$ and one for $\overline\nu_e$, no longer correlating neutrinos and antineutrinos. Therefore, the collision term for neutrinos reduces to
\begin{equation}
   {\sC}(\rho_\bp)=\frac{1}{2}\left\{\sP_E,1-\rho_\bp\right\}-\frac{1}{2}\left\{\sA_E,\rho_\bp\right\}=\sP_E-\frac{1}{2}\left\{\sA_E+\sP_E,\rho_\bp\right\},
\end{equation}
which depends only on $E=|\bp|$ and we show the $\bp$ dependence of $\rho$ as a subscript. Here, $\sP_E=Q_E \sB$ denotes the production rate of neutrinos with energy $E$, with $Q_E$ being the production rate of $\nu_e$  and $\sB$ a matrix with the only non-vanishing entry $\sB_{ee}=1$. Similarly, the net absorption matrix $\sA_E+\sP_E=2\Gamma_E \sB$ is described by the absorption rate for electron neutrinos $2\Gamma_E$. The factor $2$ is purely for notational convenience because, as we will see, flavor coherence is damped with the rate $\Gamma_E$, reflecting the usual finding that flavor coherence is damped by half the scattering rate. Analogous equations pertain to antineutrinos with the quantities $\overline{Q}_E$ and $\overline{\Gamma}_E$. The absorption and production rates are related by detailed balance as
\begin{equation}
    Q_E=\frac{2\Gamma_E}{e^{\frac{E-\mu_{\nu_e}}{T}}+1}=2\Gamma_E f^{\rm th}_{\nu_e},
\end{equation}
where $f^{\rm th}_{\nu_e}$ is the $\nu_e$ equilibrium distribution, $\mu_{\nu_e}$ their
chemical potential, and $T$ their temperature. In other words, $\mu_{\nu_e}$ and $T$ are quantities that characterize the nuclear medium and not the actual $\nu_e$ distribution. Analogous expressions pertain to $\overline\nu_e$ with $\mu_{\overline\nu_e}=-\mu_{\nu_e}$.

\subsection{Linearization in the two-flavor limit}

These general equations describe the full evolution of the neutrino plasma. However, our focus is more limited: we are primarily interested in its stability properties. Consequently, it is sufficient to consider the linear regime, where flavor correlations—represented by the off-diagonal elements of the density matrices—are small. In this regime, we can treat the diagonal elements of $\rho_\bp$ as fixed to their initial values, and study the stability of perturbations around this background.

From here on, we restrict our discussion to a two-flavor framework consisting of $\nu_e$ and $\nu_\mu$, neglecting the $\nu_\tau$ component. Following previous notation, we write the density matrices in the form
\begin{equation}
    \rho_\bp=\frac{n_\bp}{2}\,\mathbbm{1}
    +\frac{1}{2}\begin{pmatrix}G_\bp&\psi_\bp^*\\\psi_\bp&-G_\bp\end{pmatrix},
\end{equation}
where we define $\psi = 2\rho_{\mu e}$ as the measure of coherence between the two flavors. The perturbations captured by $\psi$ correspond to so-called flavor waves, whose quanta we named flavomons \cite{Fiorillo:2025npi}.

On the other hand, perturbations in the diagonal elements correspond to density waves in the plasma of $\nu_e$ and $\nu_\mu$. Their quanta are generally referred to as plasmons, associated with the total neutrino number density, given by $n=\rho_{ee}+\rho_{\mu\mu}$. In addition, there exist anti-correlated oscillations between $\nu_e$ and $\nu_\mu$ that preserve the total number density $n$. These are characterized by the variable $G=f_e-f_\mu$ and were called neutrino-plasmons~\cite{Fiorillo:2025npi}. It is straightforward to verify that the perturbations associated with plasmons and neutrino plasmons are of higher order than the flavomon perturbations described by $\psi$, and can therefore be neglected in the linear regime. Crucially, the flavor dynamics depends entirely on the difference between electron and muon neutrinos. We will refer to this quantity as the difference in lepton number ($\eln$).

In the two-flavor case, the mass matrix is parameterized in the following form~\cite{Fiorillo:2024pns}, after dropping a piece proportional to the unit matrix,
\begin{equation}
    {\sf H}_{\rm V}=\frac{\sM^2}{2E}=\frac{\omegaE}{2}
    \begin{pmatrix}-\cos2\theta_{\rm V}& \sin2\theta_{\rm V}\\
    \sin2\theta_{\rm V}&\cos2\theta_{\rm V}\end{pmatrix},
\quad\mathrm{where}\quad
\omegaE=\left|\frac{\delta m_\nu^2}{2E}\right|,
\end{equation}
$\delta m_\nu^2$ is the squared mass splitting, and $\tV$ is the mixing angle. Since the vacuum oscillation frequency $\omegaE$ is defined to be positive, $\cos2\theta_{\rm V}>0$ implies normal mass ordering, while $\cos2\theta_{\rm V}<0$ implies inverted ordering. We also introduce the notation $\tomegaE=\omega_E\cos2\tV$ for the on-diagonal energy splitting induced by the vacuum term.

We can finally extract from Eq.~\eqref{eq:QKE} the linearized equation, that is, an equations of motion for the off-diagonal element $\psi$
\begin{eqnarray}
    (\partial_t+\bv\cdot \boldsymbol{\partial}_\br)\psi_\bp&=&
    -i G_\bp \omega_E \sin(2\tV)+i\lambda \psi_\bp-i\tomegaE\psi_\bp-\Gamma_E\psi_\bp\\[1.5ex] \nonumber 
    &&{}+i\sqrt{2}\GF \int \frac{d^3\bp'}{(2\pi)^3}\left[(G_{\bp'}-\overline{G}_{\bp'})\psi_\bp-(\psi_{\bp'}-\overline{\psi}_{\bp'})G_\bp\right](1-\bv'\cdot \bv),
\end{eqnarray}
where we have parameterized matter refraction by $\sqrt{2}\GF \sN=\lambda \sB$, assuming that the dominant potential is driven by electrons. For compactness we have introduced the notation $\psi_\bp=\psi(\bp,\br,t)$, and similarly for $G_\bp$. The first term provides a perturbation transverse to the flavor direction and thus seeds instabilities if the system begins in flavor eigenstates. We also now see the reason for the factor $2$ in the definition of the scattering rate; with this choice, $\Gamma_E$ represents the rate of damping of flavor coherence. An analogous equation pertains to $\overline\psi$ for antineutrinos, where the terms proportional to $\omega_E$ and $\tomegaE$ change sign.

Finally, to unify notation with previous works, we explicitly extract the total particle density from the definitions of $G_\bp$ and $\psi_\bp$, such that $G_\bp\to G_\bp (n_\nu+n_{\overline{\nu}})$ and $\psi_\bp\to \psi_\bp (n_\nu+n_{\overline{\nu}})$, and analogous modifications for antineutrinos. Moreover, we introduce the four-dimensional notation $v^\mu=(1,\bv)$ and the integrated four-fluxes
\begin{equation}
    G^\mu=\int \frac{d^3\bp}{(2\pi)^3}\,G_\bp v^\mu
    \quad\mathrm{and}\quad
    \psi^\mu=\int \frac{d^3\bp}{(2\pi)^3}\,\psi_\bp v^\mu
\end{equation}
and similarly for $\overline{G}^\mu$ and $\overline{\psi}^\mu$. In addition, we introduce the lepton number differences
\begin{equation}
     D^\mu=G^\mu-\overline{G}^\mu
    \quad\mathrm{and}\quad
    \Psi^\mu=\psi^\mu-\overline{\psi}^\mu.
\end{equation}
We also use a general notation for their temporal and spatial components $D^\mu=(D^0,\bD)$ and $\Psi^\mu=(\Psi^0,\boldsymbol{\Psi})$. We are finally led to the compact equations
\begin{equation}\label{eq:evolution_psi}
    (\partial_t+\bv\cdot\boldsymbol{\partial}_\br)\psi_\bp=i(\lambda-\tomegaE+i\Gamma_E)\psi_\bp+i\mu\left[D\cdot v\, \psi_\bp-\Psi\cdot v\, G_\bp\right]-iG_\bp \omega_E \sin(2\tV),
\end{equation}
where the scalar four-product is understood, for example, as $D\cdot v=D_\mu v^\mu$. With our normalizations, the scale for neutrino-neutrino refraction is $\mu=\sqrt{2}\GF(n_\nu+n_{\overline\nu})$.

For completeness, we also provide the evolution equations for the occupation numbers, even though they are not relevant in the linear regime,
\begin{eqnarray}
    (\partial_t+\bv\cdot\boldsymbol{\partial}_\br)f_{e,\bp}&=&-\frac{i\omega_E \sin2\tV}{4}(\psi_\bp-\psi^*_\bp)+\frac{i \mu}{4}\left[\Psi\cdot v\, \psi^*_\bp-\Psi^*\cdot v \,\psi_\bp\right]+Q_E-2\Gamma_E f_{e,\bp},
    \nonumber\\  
    (\partial_t+\bv\cdot\boldsymbol{\partial}_\br)f_{\mu,\bp} &=&+\frac{i\omega_E \sin2\tV}{4}(\psi_\bp-\psi^*_\bp)-\frac{i \mu}{4}\left[\Psi\cdot v\, \psi^*_\bp-\Psi^*\cdot v \,\psi_\bp\right].
\end{eqnarray}
Only the $\nu_e$ occupation number has a source and sink term by our simplifying assumptions about the collision term.

Returning to the equations for $\psi$, the terms proportional to $\sin2\tV$ act only as seeds for the possible unstable modes~\cite{Fiorillo:2024pns} and are therefore neglected to determine the collective eigenmodes. These are thus determined from Eq.~\eqref{eq:evolution_psi} without the last term, and we may seek them in the form of Fourier modes $\psi_\bp\propto e^{-i\Omega t+i\bK\cdot \br}$. This ansatz is of course valid only for modes with a wavelength and period much shorter than the spatial and temporal scale of variation of the medium. Introducing the four-vector $K^\mu=(\Omega, \bK)$, and the shifted four-vector $k^\mu=(\Omega+\sqrt{2}\GF\mu D_0+\lambda,\bK+\sqrt{2}\GF\mu \bD)$, we find
\begin{equation}\label{eq:2.12}
    \psi_\bp=\frac{\mu G_\bp \Psi\cdot v}{k\cdot v-\tomegaE+i\Gamma_E}
    \quad\mathrm{and}\quad
    \opsi_\bp=\frac{\mu \oG_\bp \Psi\cdot v}{k\cdot v+\tomegaE+i\overline{\Gamma}_E}.
\end{equation}
Taking the difference among neutrinos and antineutrinos, and integrating over momentum, we are finally led to the consistency relation for the field $\Psi$
\begin{equation}
    \varepsilon_{\mu\nu}\Psi^\nu=0,
\end{equation}
where
\begin{equation}\label{eq:2.14}
    \varepsilon_{\mu\nu}=g_{\mu\nu}-\mu\int \frac{d^3\bp}{(2\pi)^3}\left[\frac{G_\bp v_\mu v_\nu}{k\cdot v-\tomegaE+i\Gamma_E}-\frac{\oG_\bp v_\mu v_\nu}{k\cdot v+\tomegaE+i\overline{\Gamma}_E}\right]
\end{equation}
is what we have called the \textit{flavor dielectric tensor} \cite{Fiorillo:2024bzm, Fiorillo:2025npi}.

The definition of this tensor is straightforward for $\mathrm{Im}(\omega)>0$, in which case the denominators in the integral never vanish. For $\mathrm{Im}(\omega)\leq0$, the integral should be defined by continuity, i.e., the integration contour in $\bp$ should be deformed such as to pass always below the values at which the denominator would vanish, a prescription that descends from causality, and has been elaborated on in much greater detail in Ref.~\cite{Fiorillo:2024bzm}. This prescription is often symbolically denoted by adding a small positive imaginary part to the eigenfrequency $\omega\to\omega+i\epsilon$, indicating that the function must be defined as the analytic continuation of the integrals for $\mathrm{Im}(\omega)>0$.

\subsection{Dispersion relation for axisymmetric distribution}

Throughout this work, we limit our investigation to axisymmetric angular distributions. Consequently, $G_\bp$ and $\oG_\bp$ depend only on the particle energy $E$ and on the angle between the neutrino velocity and the symmetry axis, $v=\bv\cdot\bn$. Additionally, we consider only modes directed along the symmetry axis, such that $k\cdot v=\omega-k v$, with $k=|\bk|$. While this is a restrictive choice, we will discuss in Sec.~\ref{sec:discussion} which of our results can be generalized to arbitrary angular distributions. Accordingly, we introduce neutrino distributions that are already integrated over the azimuthal angle, such that
\begin{equation}
    \int \frac{d^3\bp}{(2\pi)^3}\,G_\bp=\int dE\, dv\, G(v,E)
    \quad\mathrm{and}\quad
    G(v,E)=G_\bp \frac{E^2}{4\pi^2}.
\end{equation}
The flavor dielectric tensor in this very symmetric case takes the form
\begin{equation}
    \varepsilon^\mu_\nu=\begin{pmatrix}
        1-\mu I_0 & \mu I_1 & 0 & 0\\
        \mu I_1 & -1-\mu I_2 & 0 & 0\\
        0 & 0 & -1-\frac{\mu}{2}(I_0-I_2) & 0\\
        0 & 0 & 0 &-1-\frac{\mu}{2}(I_0-I_2)
    \end{pmatrix},
\end{equation}
where we use the integrals
\begin{equation}\label{eq:integrals}
    I_n=\int \frac{G(v,E) v^n dv dE}{\omega-kv-\tomegaE+i\Gamma_E+i\epsilon}-\int \frac{\oG(v,E) v^n dv dE}{\omega-kv+\tomegaE+i\overline{\Gamma}_E+i\epsilon},
\end{equation}
including the Landau prescription in the denominator as explained earlier.

As discussed many times, this system permits two degenerate axial-breaking modes, with a polarization vector $\Psi^\mu$ transverse to the axis of symmetry, satisfying the dispersion relation
\begin{equation}
    \mu(I_0-I_2)+2=0.
\end{equation}
In addition, there are two longitudinal modes, with no component transverse to the axis of symmetry, satisfying the dispersion relation
\begin{equation}\label{eq:general_dispersion_relation}
    (\mu I_0-1)(\mu I_2+1)-\mu^2 I_1^2=0.
\end{equation}
We will primarily focus on the longitudinal modes, which have been more often studied in the literature due to their full axial symmetry. However, their main properties also apply to the axial symmetry-breaking modes, as we will briefly discuss in Sec.~\ref{sec:discussion}.

For clarity, we will always assume that both $G(v,E)$ and $\oG(v,E)>0$, i.e., a dominance of $\nu_e$ and $\overline\nu_e$  over the other flavors. While generalizing our study to other scenarios is straightforward, focusing on this specific setup -- arguably the most phenomenologically relevant -- avoids repeated discussions of when this condition is necessary.

We emphasize once again, as on many previous occasions \cite{Fiorillo:2023mze, Fiorillo:2024bzm, Fiorillo:2024uki, Fiorillo:2024pns, Fiorillo:2025ank}, that the integrals over $v$ and $E$ need to be defined in a continuous manner when $\mathrm{Im}(\omega)+\Gamma_E$ passes through zero.  Causality therefore requires that for $\mathrm{Im}(\omega)+\Gamma_E>0$, the integrals are defined by Eq.~\eqref{eq:integrals}, while for $\mathrm{Im}(\omega)+\Gamma_E<0$ they are defined via its analytical continuation. In this case, the integration path in $v$ must be deformed in the complex plane to pass below the singular point where the denominator vanishes. The $i\epsilon$ prescription, made explicit in Eq.~\eqref{eq:integrals}, serves as a reminder of this analytical continuation. It means that if $\mathrm{Im}(\omega)+\Gamma_E=0$, the integral should be evaluated as the limit of a vanishingly small positive imaginary frequency $\epsilon$.

\subsection{Classification of solutions}

\subsubsection{Fast, slow, and collisional instabilities}

The solutions of this dispersion relation may involve exponentially growing modes. It is conventional to separate different classes of instabilities according to the parameters that are kept in the treatment. In particular, neglecting both $\tomegaE$ and $\Gamma_E$ leads to the \textit{fast flavor} evolution with a time scale of the order of $\mu \epsilon$. Here the parameter $\epsilon=(n_\nu-n_{\overline\nu})/(n_\nu+n_{\overline\nu})$ represents the neutrino-antineutrino asymmetry as discussed in Sec.~\ref{sec:epsilon} below. If $\tomegaE$ is kept while $\Gamma_E$ is neglected, modes that were previously stable can turn unstable, the \textit{slow modes}, with a typical growth rate $\tomegaE/\epsilon$, except for special cases where the growth rate can be as large as $\sqrt{\mu\tomegaE}$. This fast vs.\ slow terminology is mostly historically motivated because it was thought that the scale $\sqrt{\mu\tomegaE}\ll\mu$ was generic, but today we know that the slow modes can actually have faster growth rates than the fast ones \cite{Fiorillo:2024pns}. The reason for these scalings will be reviewed later, in Sec.~\ref{sec:superluminal}.  If $\Gamma_E$ is kept while $\tomegaE$ is neglected, modes that are fast-stable can become unstable, the \textit{collisional instabilities}. If $\Gamma_E$ and $\tomegaE$ are both included, either can trigger an instability of fast-stable modes, which can therefore become slow-unstable or collisional-unstable, depending on which scale is the fastest.

\subsubsection{Subluminal vs.\ superluminal modes}

However, to understand the physical mechanism of instability and to obtain meaningful approximations for the unstable modes, a different classification proves more useful: one based on whether the phase velocity $\omega/k$ is superluminal, subluminal, or near-luminal. This separation is motivated by the physical fact that superluminal modes cannot move in phase with any neutrino in the plasma. Such modes can only gain energy from neutrinos in a non-resonant manner -- meaning that no particular neutrino energy or direction preferentially transfers energy to the mode. For subluminal modes, the opposite holds, especially in the case of weak instabilities, where neutrinos moving in phase with the wave can resonantly supply it with energy.

In the classical wave picture, this occurs because neutrinos in phase with the wave can coherently exchange energy, as they remain locked in phase with it. In the quantum picture, flavomons moving in phase with neutrinos can be produced on-shell via neutrino decay. Due to this clear physical distinction, we will discuss these classes separately.

\subsubsection{Gapped vs.\ gapless modes}

A general property of the dispersion relation Eq.~\eqref{eq:general_dispersion_relation} is that the integrals $I_n$ appear multiplied by the factor $\mu$, which is large compared to all other scales set by masses and collisions. Therefore, to satisfy the dispersion relation, which contains an inhomogeneous polynomial, it would appear necessary that the oscillation frequency $\mathrm{Re}(\omega)$ of a solution is of the order of $\mu$, so that the products $\mu I_n$ become factors of order unity. 

Therefore, we will initially consider only cases where $\mathrm{Re}(\omega) \gg \tomegaE$ and $\Gamma_E$. We refer to these solutions as \textit{gapped modes}, since their frequency $\omega$ is always characterized by the scale $\mu$ and thus remains cleanly separated from zero.

Later, we will discuss the possibility of collective modes with $\mathrm{Re}(\omega) \sim \tomegaE$ or $\Gamma_E$. These will be referred to as \textit{gapless modes}. While they are not strictly gapless -- in the sense of having zero real eigenfrequency -- their typical frequencies are much smaller than those of the gapped modes. In the limit of vanishing vacuum frequencies and collisional rates, they tend toward unphysical zero-frequency modes.  Our terminology is borrowed from condensed matter physics, where a mode is said to be gapless if there exists a wave number $k$ for which $\omega$ vanishes.

\section{Scales for order-of-magnitude estimates}\label{sec:scales}

One of our main goals is to understand the typical orders of magnitude for the growth rates of unstable modes across different wavenumbers and environments. For such estimates, it is useful to rely on a few indicative energy scales that characterize the problem. To express hierarchies of scales, we will often write inequalities of the type $\mu\epsilon\gg\tomegaE$, which are understood to mean  $|\mu\epsilon|\gg|\tomegaE|$, as a way to avoid cluttering the expressions.

\subsection{Refractive energy scale}

The energy shift caused by neutrino-neutrino refraction is represented by a single number, the refractive energy scale $\mu=\sqrt{2}\GF (n_\nu+n_{\overline{\nu}})$. In the decoupling region for SN neutrinos, their typical density is in the ballpark of $10^{32-34}$~cm$^{-3}$, at the lower end corresponding to
$\mu\sim 10\; \mu\mathrm{eV}\sim 1\;\mathrm{cm}^{-1}$. 

A first simple but important insight is that this scale is much smaller than a typical neutrino energy $E\sim 10$~MeV, indicating a strong hierarchy between the neutrino kinetic energy reservoir and their interaction energy. Even in the free-streaming regime, where neutrino-neutrino collisions are negligible, refraction still allows for energy exchange in an amount of order $\mu$ \cite{Fiorillo:2024fnl}. Although $\mu\ll E$ implies that the relative change in kinetic energy is extremely small, the refractive energy itself changes significantly. Consequently, conservation of neutrino-neutrino interaction energy does not hold, due to the effectively infinite reservoir of kinetic energy.

\subsection{Lepton asymmetry}
\label{sec:epsilon}

While the parameter $\mu$ is commonly used as the sole measure of interaction strength, additional specifications may actually be necessary. In the so-called fast limit, in which neutrino masses and collisions are neglected, neutrino-neutrino refraction depends not on the total neutrino number density, but rather on the neutrino lepton number density, i.e., on $n_\nu-n_{\overline{\nu}}$. More precisely, this should be quantified by the so-called $e$-$x$-lepton-number density $(n_{\nu_e}-n_{\overline\nu_e})-(n_{\nu_x}-n_{\overline\nu_x})$, where $x$ stands for $\mu$ or $\tau$. However, in a SN environment, the heavy-flavor states are strongly subdominant compared to the electron-flavored ones. Therefore, we can often approximate the lepton number asymmetry using a single parameter, $\epsilon=(n_\nu-n_{\overline{\nu}})/(n_\nu+n_{\overline{\nu}})$.

In the fast limit, the typical energy scale associated with the wavelength and period of collective neutrino modes is therefore $(\mu \epsilon)^{-1}$.
It is important to note that, for specific energy and angle distributions, the interpretation of $\epsilon$ becomes inherently ambiguous, as the $\nu$ and $\overline\nu$ distributions may differ significantly at some energies and angles while being nearly identical at others. Still, we assume that if $\epsilon\sim 1$ (large $\nu$--$\overline\nu$ differences), this probably holds across the entire distribution, except perhaps in localized regions such as angular crossings. Conversely, when $\epsilon\ll 1$, corresponding to nearly equal numbers of neutrinos and antineutrinos, $\epsilon$ likely remains small across all energies and angles.

\subsection{Masses and vacuum oscillation frequency}
\label{sec:masses}

Neutrino masses introduce yet another scale. The role of flavor mixing is limited to providing a seed for unstable modes -- in the flavomon language, for the stimulated decay into flavomons. Neutrino masses, on the other hand, are more consequential, as they induce an energy shift between neutrinos and antineutrinos. In a two-flavor setup, this difference is characterized by the frequency $\tomegaE=\delta m_\nu^2 \cos2\tV/2E$. Although $\tomegaE$ depends on energy, for order-of-magnitude estimates we adopt its value for the characteristic neutrino energy $E\sim 10$~MeV. With $\delta m_\nu^2\sim 10^{-3}$~eV$^2$ and $\cos2\tV\sim 1$, we find $\tomegaE\sim 5\times 10^{-11}\; \mathrm{eV}\sim 3\times 10^{-6}\;\mathrm{cm}^{-1}$. Whenever we refer to order-of-magnitude estimates, we implicitly assume this fixed value of $\tomegaE$, treating it as energy-independent. The vacuum frequency is therefore 5--6 orders of magnitude smaller than the neutrino-neutrino refractive energy shift. The dimensionless ratio $\tomegaE/\mu\epsilon\sim 10^{-5}$ is the first small parameter of our problem.

\subsection{Collisions}

A final scale is provided by the damping rate for flavor coherence $\Gamma_E$ (and $\overline\Gamma_E$ for $\overline\nu$). Such damping is caused by collisions and we will limit our discussion to the dominant effect provided by $\nu_e$ and $\overline\nu_e$ beta processes on nucleons. Notice that these are typically very different for $\nu_e$ and $\overline\nu_e$, a difference that is needed for collision-driven instabilities. The coherence damping rate is half the charged-current scattering rate and thus proportional to $E^2$. The scattering rate changes not only with neutrino energy, but also across the SN profile, since the density of nuclei changes. However, as a rough proxy, we may notice that in the decoupling region, lying at approximately the radius of the protoneutron star $r_{\rm NS}\sim 20$~km, the collisional rate must be roughly of the inverse order of magnitude of this distance, so $\Gamma_E\sim 5\times 10^{-7}\;\mathrm{cm}^{-1}\sim 10^{-11}\;\mathrm{eV}$.

Such estimates represent rough orders of magnitudes, and in this sense it does not matter if we refer to $\nu_e$ or $\overline\nu_e$ or the exact energy. The main point is that $\Gamma_E\ll \mu \epsilon$, but we observe a peculiar numerical coincidence, which is that in the decoupling region $\Gamma_E$ and $\tomegaE$ are relatively close to each other. This suggests that the concept of slow and collisional flavor limit, in which only vacuum frequency or only neutrino-matter collisions, respectively, are taken into account, may never be a very good one for precision predictions, although they serve as fundamental starting points for conceptual understanding. 

\section{Superluminal gapped modes}\label{sec:superluminal}

According to our new classification, we begin with superluminal modes, i.e., modes for which individual neutrinos in the flavor plasma can never be in phase with the wave. We focus on gapped modes, where the frequency never becomes as small as $\tomegaE$ or $\Gamma_E$. A characteristic property of these modes is that $k\ll \omega$, allowing us to understand their behavior by expanding the dispersion relation for small $k$. Here, we consider only the limiting case $k=0$, restricting the expansion to zeroth order; extending to higher orders does not introduce any new conceptual challenges. Recall, however, that $k=0$ does not correspond to the homogeneous case, which is instead defined by $K=0$, i.e., the vanishing of the physical wavenumber.

\subsection[Dispersion relation for vanishing \texorpdfstring{$k$}{}]{Dispersion relation for vanishing \texorpdfstring{\boldmath$k$}{}}

In the limit of vanishing wavenumber ($k=0$), the integrals of Eq.~\eqref{eq:integrals} that enter the dispersion relation are
\begin{equation}\label{eq:In-homogeneous}
    I_n=\int \frac{G_n(E) dE}{\omega-\tomegaE+i\Gamma_E}-\int \frac{\oG_n(E) dE}{\omega+\tomegaE+i\overline{\Gamma}_E},
\end{equation}
where we introduce the notation $G_n(E)=\int G(v,E) v^n dv$ and similarly for $\oG_n(E)$. Thus, for $k=0$, the dependence on the neutrino angular distribution is completely encoded in the angular moments and does not descend from local properties of the angular distribution. This key fact, which we have emphasized throughout our previous series of papers \cite{Fiorillo:2024bzm, Fiorillo:2024uki, Fiorillo:2024pns, Fiorillo:2025ank}, stems from a specific physical property: superluminal modes are non-resonant and therefore interact with the entire distribution simultaneously.

Let us now examine separately the behavior of this simplified dispersion relation in the three different regimes of fast, slow, and collisional modes.

\subsection{Fast flavor limit}

In the fast flavor limit, defined by $\tomegaE=0$ and $\Gamma_E=\overline{\Gamma}_E=0$, the integrals reduce to $I_n=D_n/\omega$, where $D_n=\int dE\left[G_n(E)-\oG_n(E)\right]$. The corresponding dispersion relation admits the solution
\begin{equation}\label{eq:fast_superluminal}
    \omega_\pm=\frac{\mu}{2}\left[D_0-D_2\pm\sqrt{(D_0+D_2)^2-4D_1^2}\right].
\end{equation}
To understand the crucial orders of magnitude we introduce the asymmetry parameter $\epsilon$ discussed in Sec.~\ref{sec:epsilon}, which is here representative of $D_n/G_n$. In the spirit of such a rough estimate, the oscillation frequency, and potentially the growth rates of fast modes, are of order $\mu \epsilon$. However, the two fast modes are generally real and therefore stable, unless $(D_0+D_2)^2-4D_1^2<0$, corresponding to a strongly unstable case. It requires a deep angular crossing because this quantity is positive for a distribution with a very shallow angular crossing. This insight connects with the general finding that weak fast instabilities are not strongly superluminal. We refer to our theory of fast flavor conversions~\cite{Fiorillo:2024bzm, Fiorillo:2024uki, Fiorillo:2024dik} for a more detailed treatment of the fast flavor limit.

\subsection{Slow flavor limit}\label{sec:slow_superluminal}

In the purely slow limit ($\tomegaE\neq 0$ while $\Gamma_E=\overline{\Gamma}_E=0$), there is no general solution for the $I_n$ integrals of Eq.~\eqref{eq:In-homogeneous}, since the denominators depend on energy. However, one can still explore limiting cases to gain further insight and in particular we assume $\tomegaE\ll\mu\epsilon$ as motivated in Sec.~\ref{sec:masses}. 
Beyond this hierarchy, we study two regimes, one of an even smaller $\tomegaE\ll \mu \epsilon^2$ and another one $\mu \epsilon^2\ll\tomegaE\ll\mu \epsilon$, which requires $\epsilon\ll1$ to exist at all.

In the fast limit we have found that $\mathrm{Re}(\omega)\sim \mu \epsilon$, implying that a small $\tomegaE$ should only spawn a small modification. In the denominator of the integrand of $I_n$, therefore $\mathrm{Re}(\omega)\gg\tomegaE$ and we can expand the integrals as
\begin{equation}\label{eq:TaylorSlow}
    I_n=\frac{D_n}{\omega}+\frac{S_n}{\omega^2},
\quad\mathrm{where}\quad
   S_n=\int \left[G_n(E)+\oG_n(E)\right]\tomegaE\,dE. 
\end{equation}
This treatment exactly parallels the one we gave for a monochromatic energy distribution~\cite{Fiorillo:2024pns} and illuminates that even for a non-monochromatic distribution the properties of these modes remain the same.

The dispersion relation has now turned completely algebraic, and the information on the energy distribution is entirely encoded in its moments. Since $\mathrm{Re}(\omega)\sim \mu \epsilon$, the first term in $I_n$ has order of magnitude $D_n/\omega\sim 1/\mu$, while the second term is $S_n/\omega^2\sim \tomegaE/\mu^2\epsilon^2$, where by $\tomegaE$ we mean a typical order of magnitude or average value over the energy distribution. The behavior is very different if $\tomegaE\ll \mu \epsilon^2$ or $\tomegaE\gg \mu \epsilon^2$. 

For $\tomegaE\ll \mu \epsilon^2$, the terms containing $S_n$ in $I_n$ can to a first approximation be neglected altogether and the slow modes are only small corrections of the fast ones. The precise values of the eigenfrequencies can be obtained by a Taylor expansion in terms of the small terms containing $S_n$, obtaining
\begin{equation}
    \frac{(\mu D_0-\omega)(\mu D_2+\omega)-\mu^2 D_1^2}{\omega^2}+\mu\,\frac{\mu (D_2 S_0-2D_1 S_1+D_0 S_2)+S_0 \omega-S_2 \omega}{\omega^3}\simeq 0.
\end{equation}
The limit $S_n\to 0$ returns the original fast eigenfrequencies in Eq.~\eqref{eq:fast_superluminal}, and the corrections can be found by treating the second term perturbatively. The resulting expressions are lengthy and not particularly illuminating, except for one feature: for a shallow angular crossing, the corrections to the eigenfrequency are also real. Therefore, the vacuum frequency cannot produce slow superluminal instabilities, provided $\tomegaE\ll \mu \epsilon^2$.

In the opposite regime $\tomegaE\gg \mu \epsilon^2$, we can make the opposite choice and neglect all the terms containing $D_n$. There is no contradiction here
with our original Taylor expansion Eq.~\eqref{eq:TaylorSlow} which requires $\tomegaE\ll \mu \epsilon$, and the fact that the first-order term $S_n/\omega^2$ may dominate over the zero-order one $D_n/\omega$ if $\tomegaE\gtrsim \mu \epsilon^2$. For each energy, the Taylor expansion consistently produces higher-order terms which are smaller, but when integrating over all energies and over neutrinos and antineutrinos, the zero-order terms suffer a near-cancellation. Thus, the solutions of the dispersion relation become $\omega_{1,\pm}=\omega_{\pm}^{\rm slow}$ and $\omega_{2,\pm}=-\omega_{\pm}^{\rm slow}$, with
\begin{equation}\label{eq:slow_superluminal}
    \omega^{\rm slow}_{\pm}=\sqrt{\mu \frac{S_0-S_2\pm\sqrt{(S_0+S_2)^2-4S_1^2}}{2}}.
\end{equation}
These quantities can be complex, so in the regime $\tomegaE\gg \mu \epsilon^2$ the vacuum frequency can trigger slow superluminal instabilities. In fact, we immediately recognize the appearance of the slow pendulum mode: for an isotropic distribution ($S_1=0$), the two solutions decouple in $\omega^{\rm slow}_+=\sqrt{S_0}$ and $\omega^{\rm slow}_-=i\sqrt{S_2}$. 

These correspond to the isotropic and anisotropic instabilities identified first in Ref.~\cite{Raffelt:2007yz}, and then discussed in general terms in our theory of slow flavor conversions~\cite{Fiorillo:2024pns, Fiorillo:2025ank}. The isotropic mode may be denoted as a monopole mode; it is unstable when $S_0< 0$, and the corresponding perturbation is isotropic (its eigenmode is $\Psi^0\neq 0$ and $\boldsymbol{\Psi}=0$). The anisotropic mode may be denoted as a dipole mode; it is unstable when $S_2>0$, and its perturbation is anisotropic, with $\Psi^0=0$ and $\boldsymbol{\Psi}\neq 0$, identifying a privileged direction and a dipolar perturbation $\psi_\bp\propto \bv \cdot \boldsymbol{\Psi}$. For normal ordering ($S_n>0$ if $G_n(E)>0$ and $\oG_n(E)>0$) only the dipole mode is unstable, while for inverted ordering only the monopole mode, also known as the slow pendulum, is unstable. Equation~\eqref{eq:slow_superluminal} generalizes this result to an arbitrary angular distribution, and shows clearly the conditions for a slow superluminal instability. The most important takeaway is that in this case the typical growth rate is of the order of $\omega^{\rm slow}_\pm\sim \sqrt{\mu \tomegaE}\gg \mu \epsilon$, so we are still in a large oscillation frequency regime.

\subsection{Collisional flavor limit}\label{sec:superluminal_collisional}

The treatment of the collisional flavor limit mirrors the slow case. As we will see, all of the properties of collisional modes that have been discussed in the literature can be understood by the same methods. Once again, we focus on modes with $\mathrm{Re}(\omega)\sim \mu \epsilon$. Thus, we can once again expand the integrals
\begin{equation}
    I_n=\frac{D_n}{\omega}-\frac{i\Sigma_n}{\omega^2},
    \quad\mathrm{where}\quad
    \Sigma_n=\int \left[G_n(E)\Gamma_E-\oG_n(E) \overline{\Gamma}_E\right] dE.
\end{equation}
The energy integrals $\Sigma_n$ now involve $\Gamma_E$ instead of $\tomegaE$ in the earlier $S_n$. Their order of magnitude is generically $\Sigma_n\sim \Gamma_E$, although it can be smaller for fine-tuned conditions when $G_n(E)\Gamma_E\simeq \overline{G}_n(E) \overline{\Gamma}_E$. Therefore, we will write the order of magnitude of $\Sigma_n\sim \Gamma_E \epsilon_\Gamma$ in terms of a new asymmetry parameter $\epsilon_\Gamma$ that quantifies the asymmetry between $G_n(E)\Gamma_E$ and $\oG_n(E) \overline{\Gamma}_E$. As in the slow limit, we can separate two regimes, according to which term is the largest in the integrals $I_n$, and depends on whether $\Gamma_E\ll \mu \epsilon^2/\epsilon_\Gamma$ or $\Gamma_E\gg \mu \epsilon^2/\epsilon_\Gamma$.

In the first regime, defined by $\Gamma_E\ll \mu \epsilon^2/\epsilon_\Gamma$, we  expand the dispersion relation to first order in $\Sigma_n$
\begin{equation}\label{eq:approximate_dispersion_collisional}
    \frac{(\mu D_0-\omega)(\mu D_2+\omega)-\mu^2 D_1^2}{\omega^2}-i\mu\,\frac{\mu (D_2 \Sigma_0-2D_1 \Sigma_1+D_0 \Sigma_2)+\Sigma_0 \omega-\Sigma_2 \omega}{\omega^3}\simeq 0.
\end{equation}
To zeroth order in $\Sigma_n$ we recover the usual fast modes of Eq.~\eqref{eq:fast_superluminal}. In contrast to slow modes, however, the small correction of order $\Gamma_E$ to the eigenvalues, that is obtained by treating the terms including $\Sigma_n$ perturbatively in Eq.~\eqref{eq:approximate_dispersion_collisional}, are not real. Therefore, even for $\Gamma_E\ll \mu \epsilon^2$, the modes can be damped or unstable. The corrections to the eigenvalues can be found explicitly, but the expressions are quite lengthy. 

Instead, we focus on the special case of an isotropic distribution, implying $D_1=\Sigma_1=0$, and the corrections to the two eigenmodes are found to be
\begin{equation}\label{eq:branches_collisional_superluminal}
    \omega_+\simeq \mu D_0-i\frac{\Sigma_0}{D_0}
    \quad\mathrm{and}\quad
    \omega_-\simeq \mu D_2-i\frac{\Sigma_2}{D_2}.
\end{equation}
Thus, the previously real-valued modes have acquired an imaginary eigenfrequency, that corresponds to damping if $\Sigma_0>0$ (notice that $\Sigma_2=\Sigma_0/3$ for an isotropic distribution), while it corresponds to growth if $\Sigma_0<0$. 

The unstable monopole mode corresponds here to the collisional instability first identified in Ref.~\cite{Johns:2021qby}, see also Ref.~\cite{Xiong:2022zqz}, and we later dubbed it the first branch of collisional instability~\cite{Fiorillo:2023ajs}. The quadrupole mode $\omega_-$, instead, was discussed in Ref.~\cite{Liu:2023pjw}, and breaks isotropy. This behavior appears completely natural from our unifying perspective on the dispersion relation. In particular, the perturbative solution of Eq.~\eqref{eq:approximate_dispersion_collisional} provides the general eigenmodes for an arbitrary angular distribution, for which the monopole and quadrupole modes are mixed with each other.

In the second regime, $\Gamma_E\gg \mu \epsilon^2/\epsilon_\Gamma$, once more, to first approximation we may keep only the terms proportional to $\Sigma_n$ in $I_n$. Then, the solutions of the dispersion relation can be found in general, and are sufficiently simple that we can write them explicitly even for an anisotropic distribution in the form $\omega_{1,\pm}=\omega^{\rm col}_\pm$ and $\omega_{2,\pm}=-\omega^{\rm col}_\pm$ with
\begin{equation}
    \omega^{\rm col}_\pm=\sqrt{\mu\frac{i(\Sigma_2-\Sigma_0)\pm \sqrt{4\Sigma_1^2-(\Sigma_0+\Sigma_2)^2}}{2}},
\end{equation}
to be compared with the analogous slow expression of Eq.~\eqref{eq:slow_superluminal}. Among these four modes, two will be growing; a collisional instability is now unavoidable, and its growth rate has an order of magnitude $\mathrm{Im}(\omega)\sim \sqrt{\Gamma_E \mu \epsilon_\Gamma}$. The enhancement of the growth rate
when $\Gamma_E\gg \mu \epsilon^2/\epsilon_\Gamma$ has been dubbed a resonance-like structure in Refs.~\cite{Liu:2023pjw,Xiong:2022zqz}. From our perspective, this enhancement is of the same nature as the square-root behavior of slow instabilities $\mathrm{Im}(\omega)\sim \sqrt{\mu \tomegaE}$, which is valid only for $\tomegaE\gg \mu\epsilon^2$.

While we have treated the slow and collisional flavor limits separately, the developed approximations are identical as they are based on expanding the denominator for $\mathrm{Re}(\omega)\gg \tomegaE$ or $\Gamma_E$. Therefore, it is straightforward to use the same methods to treat a mixed regime where $\tomegaE$ and $\Gamma_E$ are comparable.

\subsection{Summary}

Superluminal modes, in the fast regime, are usually stable, unless there is a strong angular crossing, which might turn them unstable and damped. In particular, around $k=0$, in the fast limit there is a single pair of eigenmodes. We have shown that the vacuum frequency can alter the eigenfrequency by small terms of order $\tomegaE/\epsilon$, which however are real and do not change the nature of the mode, while collisions introduce corrections of order $\Gamma_E \epsilon_\Gamma/\epsilon$, which instead are imaginary and turn the modes into either damped or growing ones, the latter only if $\Sigma_0<0$ for an isotropic gas, which requires antineutrinos to scatter more rapidly than neutrinos. We emphasize that all the results in this section refer only to modes with $\mathrm{Re}(\omega)\gg \tomegaE, \Gamma_E$, which we have dubbed gapped modes, and which exist also in the fast limit, and are therefore perturbations of fast modes.

For realistic SN environments, presumably $\Sigma_0>0$, so even in the collisional regime these modes are likely damped and, overall, superluminal gapped modes are probably not responsible for instabilities. However, as a matter of completeness, in Sec.~\ref{sec:example_distributions} we will show also examples in which $\Sigma_0<0$. 

If $\tomegaE\gg \mu \epsilon^2$, or $\Gamma_E\gg \mu \epsilon^2/\epsilon_\Gamma$, we enter a regime in which vacuum frequency and collisions dominate the dispersion relation, so the eigenvalues can become complex even in the slow limit with a typical growth rate $\mathrm{Im}(\omega)\sim \sqrt{\mu\tomegaE}$, or $\mathrm{Im}(\omega)\sim \sqrt{\Gamma_E \mu \epsilon_\Gamma}$. The slow, monopole mode for an isotropic distribution is the well-known slow flavor pendulum; our generalized perspective shows that it is only a part of a wider family of unstable superluminal modes in the slow and collisional regime. Notice that, for collisional modes, this regime has often been dubbed resonance-like, because it leads to a huge enhancement of the growth rate\footnote{Ironically, in the analogous case of slow modes, the growth rate of the order of $\sqrt{\mu \tomegaE}$ was perceived as small, compared with the fast growth rate, and was therefore the main reason for the ``slow'' terminology. However, even in this case, the growth rate of order $\sqrt{\mu\tomegaE}$ is hugely enhanced compared to a typical regime, in which the growth rate is of order $\tomegaE/\epsilon$.}, which is now of the order of $\sqrt{\Gamma_E \mu \epsilon_\Gamma}$ rather than $\Gamma_E\epsilon_\Gamma/\epsilon$. We stress that this particular connotation resonance is entirely different from the notion of wave-particle resonance that drives the growth of flavor waves in the fast and slow regime as we have discussed in many papers~\cite{Fiorillo:2024bzm, Fiorillo:2024uki, Fiorillo:2024pns, Fiorillo:2025ank, Fiorillo:2024dik, Fiorillo:2025npi}.

Our main result of this section was to provide an approximate dispersion relation for superluminal modes that is completely algebraic, depending only on moments of angular and energy distributions. We have limited our treatment to $k=0$, but for $k\ll \omega$ it is straightforward
to obtain small-$k$ corrections by Taylor-expanding the denominators in the integrals $I_n$. Both for collisional and slow modes, these algebraic relations mostly existed already scattered across various papers. Our unified treatment shows that the simplicity of the algebraic treatment fundamentally descends from the superluminal nature of the modes, whose phase evolves much more rapidly than that of individual neutrinos and therefore do not single out any specific neutrino velocity. Therefore, only the moments of the angular distribution are relevant to determine the dispersion relation.

\section{Subluminal gapped modes}\label{sec:subluminal}

We now turn to subluminal modes, defined by $\omega < k$, which are characterized by the feature that the real part of the denominators in the integrals $I_n$ can vanish for certain neutrino velocities and directions. We have dubbed such modes resonant \cite{Fiorillo:2024bzm, Fiorillo:2024uki}, because the corresponding neutrinos move in phase with the wave and can thus be strongly affected by their interaction through the locked phase relation. This situation parallels the standard resonant interaction of electrons with plasma waves. We discuss the fast, slow, and collisional flavor cases separately.

\subsection{Fast flavor limit}

We have previously studied the fast flavor limit of the dispersion relation given in Eq.~\eqref{eq:general_dispersion_relation} in great detail \cite{Fiorillo:2024bzm, Fiorillo:2024uki, Fiorillo:2024dik}. The integrals over energy can be performed explicitly so that
\begin{equation}
    I_n=\int \frac{D(v) v^n dv}{\omega-kv+i\epsilon}
\end{equation}
after introducing the lepton-number angle distribution $D(v)=\int dE\,[G(E)-\oG(E)]$. The behavior of fast modes depends on the presence or absence of an angular crossing. 

In the absence of a crossing, $D(v)$ never vanishes, and for any $\omega<k$ we find that, if $\mathrm{Im}(\omega)=0$, the integrals $I_n$ are always complex, since
\begin{equation}
    I_n=\fint \frac{D(v) v^n dv}{\omega-kv}-i\pi \int \delta(\omega-kv) D(v) v^n dv.
\end{equation}
Assuming a real-valued subluminal mode is then self-contradictory, unless the imaginary parts of different $I_n$ cancel. (Under certain circumstances, this may occur for the fast flavor pendulum due to its Goldstone nature that originates from translational invariance~\cite{Fiorillo:2024dik}.) Likewise, lepton-number conservation prevents an instability without a crossing \cite{Johns:2024bob, Fiorillo:2024bzm}. In general, $\mathrm{Im}(\omega)<0$, corresponding to \textit{Landau damping}, is thus a generic effect, stemming from the $i\epsilon$ prescription in Eq.~\eqref{eq:integrals}, as we were first to recognize \cite{Fiorillo:2024bzm, Fiorillo:2024uki}.

The presence of an angular crossing, in contrast, means that $D(v_{\rm cr})=0$ for a certain velocity $v_{\rm cr}$, implying that for $\omega=v_{\rm cr} k$ the integrals $I_n$ are purely real. Consequently, there exists a pair of modes with $\omega=v_{\rm cr} k$ and $\mathrm{Im}(\omega)=0$. In addition, continuity implies that $\mathrm{Im}(\omega)>0$ either for $\omega>v_{\rm cr}k$ or $\omega<v_{\rm cr}k$, indicating an instability. We have previously used this property to obtain approximate expressions for the growth rates of fast unstable modes resonating with neutrinos near $v_{\rm cr}$~\cite{Fiorillo:2024uki}. In particular, parameterizing $\omega=\omega_R+i\omega_I$, with $\omega_I\ll \omega_R$ (since $\omega_I$ vanishes for $\omega_R=v_{\rm cr}k$), we find to first order
\begin{equation}
     I_n=\int \frac{D(v) v^n dv}{\omega_R-kv+i\epsilon}-i \omega_I \int\frac{D(v) v^n dv}{(\omega_R-kv+i\epsilon)^2}.
\end{equation}
This Taylor expansion allows us to express $\omega_I$ in terms of the integrals evaluated at $\omega_R$; for the resulting approximate growth rates see Ref.~\cite{Fiorillo:2024uki}.

\subsection{Slow flavor limit}

In the slow flavor limit, we may use $\tomegaE\ll \omega,k$ to perform a Taylor expansion
\begin{equation}\label{eq:expansion_subluminal_In}
    I_n=\int \frac{D(v) v^n dv}{\omega-kv+i\epsilon}-\int\frac{S(v) v^n dv}{(\omega-kv+i\epsilon)^2},
\end{equation}
where 
\begin{equation}
    S(v)=\int dE \left[G(v,E)+\oG(v,E)\right]\tomegaE.
\end{equation}
The situation is similar to Sec.~\ref{sec:slow_superluminal}, that is,
if $\tomegaE\ll \mu \epsilon^2$, the terms proportional to $S(v)$ are a small correction to the unperturbed fast modes, whereas if $\tomegaE\gg \mu \epsilon^2$, it is only these terms that dominate the integrals. Let us consider separately these two cases.

For $\tomegaE\ll \mu \epsilon^2$ we first neglect the terms proportional to $S(v)$, implying that the slow modes are only a small perturbation of the fast ones. If there is no angular crossing, all unperturbed eigenmodes are Landau-damped with a typical rate $\mu \epsilon$. Therefore, $\tomegaE$ would at most slightly renormalize the oscillation frequency and growth rate by terms of order $\tomegaE\ll \mu \epsilon$. Therefore, we do not expect subluminal instabilities, a conclusion that we had previously reached numerically~\cite{Fiorillo:2024pns, Fiorillo:2025ank}.

If there is an angular crossing, for $\omega=v_{\rm cr}k$ the imaginary parts of $I_n$ vanish in the fast flavor limit, whereas Eq.~\eqref{eq:expansion_subluminal_In} shows that the imaginary parts of the terms proportional to $S(v)$ do \textit{not} vanish. Therefore, close to $v_{\rm cr}$, the vacuum frequency can renormalize the previously stable mode and endow it with a small imaginary part. Expanding Eq.~\eqref{eq:expansion_subluminal_In}
\begin{equation}
    I_n=\int \frac{D(v) v^n dv}{\omega_R -kv+i\epsilon}-i\omega_I \int \frac{D(v) v^n dv}{(\omega_R-kv+i\epsilon)^2}-\int \frac{S(v) v^n dv}{(\omega_R-kv+i\epsilon)^2}
\end{equation}
reveals that the imaginary part induced by the vacuum frequency will be of the order of $\omega_I\sim \tomegaE/\epsilon$, since $S(v)\sim \tomegaE$ while $D(v)\sim \epsilon$. Therefore, the modes at $\mathrm{Re}(\omega)=v_{\rm cr} k$, which in the fast limit are purely real, now acquire an imaginary part, whose order of magnitude is $\mathrm{Im}(\omega)\sim \tomegaE/\epsilon$. A slight renormalization of the dispersion relation is significant only close to $v_{\rm cr}$, whereas away from it the effect of $\tomegaE$ is generically small. 

In the opposite regime, $\tomegaE\gg \mu \epsilon^2$, to first approximation we keep only the second term in Eq.~\eqref{eq:expansion_subluminal_In}. Now $I_n$ is no longer of the order of magnitude $\epsilon/\omega$, but rather $I_n\sim \tomegaE/\omega^2$. For the dispersion relation to be satisfied, which requires $\mu I_n \sim 1$, the eigenmodes must now have frequencies of the order of $\omega\sim \sqrt{\tomegaE \mu}$. We conclude that even in the subluminal regime, the square-root scaling that was often advocated for slow modes only holds for a fine-tuned equality between neutrinos and antineutrinos.

\subsection{Collisional flavor limit}

As in the superluminal case, with $\tomegaE\to i \Gamma_E$ the collisional limit is a straightforward extension of the slow case. Assuming again $\Gamma_E\ll \omega$ and $k$, we expand the integrals as
\begin{equation}\label{eq:collisional_subluminal_expansion}
    I_n=\int \frac{D(v) v^n dv}{\omega-kv+i\epsilon}-i\int \frac{\Sigma(v) v^n dv}{(\omega-kv+i\epsilon)^2},
\end{equation}
with
\begin{equation}
    \Sigma(v)=\int dE\left[G(v,E)\Gamma_E-\oG(v,E)\overline{\Gamma}_E\right].
\end{equation}
The subsequent discussion parallels entirely the one for the slow flavor limit.

For $\Gamma_E\ll \mu \epsilon^2/\epsilon_\Gamma$, once more the terms proportional to $\Sigma(v)$ can be neglected to first approximation. Collisions slightly perturb the frequencies of the fast subluminal modes, which we know to be Landau-damped if there is no angular crossing, or otherwise transition from Landau-damped to unstable. Therefore, just as in the slow limit, collisions can either slightly alter the damping rate in the absence of a crossing, without qualitative consequence, or otherwise can slightly change the growth and damping rates for modes resonant with neutrinos around $v_{\rm cr}$. Either way, collisions do not have a qualitatively large impact.

For $\Gamma_E\gg \mu \epsilon^2/\epsilon_\Gamma$, instead we may keep only the terms depending on $\Sigma(v)$ in Eq.~\eqref{eq:collisional_subluminal_expansion}. Without repeating the identical discussion of the slow limit, we conclude that the growth or damping rate of subluminal modes will be of the order $\mathrm{Im}(\omega)\sim \sqrt{\Gamma_E \mu \epsilon_\Gamma}$.

\subsection{Summary}

In this section, we have comprehensively studied subluminal modes in the fast regime. Without an angular crossing, they are always Landau-damped, whereas otherwise they can transition to growing modes if they resonate primarily with neutrinos on the ``flipped'' side of the crossing as we have previously detailed~\cite{Fiorillo:2024bzm, Fiorillo:2024uki, Fiorillo:2024dik}. The main innovation here was to evaluate the impact of the vacuum frequency and collisions on these modes. 

For $\tomegaE\ll \mu \epsilon^2$ (or for $\Gamma_E\ll \mu \epsilon^2/\epsilon_\Gamma$), we have found that these small perturbations simply renormalize the growth or damping rates by terms of order $\tomegaE/\epsilon$ (or $\Gamma_E\epsilon_\Gamma/\epsilon$), which is a small correction of the typical damping and growth rates $\mathrm{Im}(\omega)\sim \mu \epsilon$ of fast modes. Therefore, neutrino masses and collisions do not qualitatively change the picture. 

The situation changes if $\tomegaE\gg \mu \epsilon^2$ (or $\Gamma_E\gg \mu \epsilon^2/\epsilon_\Gamma$), because now these corrections dominate, the growth or damping rates completely change, and acquire $\mathrm{Im}(\omega)\sim \sqrt{\tomegaE \mu}$ or $\mathrm{Im}(\omega)\sim \sqrt{\Gamma_E \mu \epsilon_\Gamma}$ by order of magnitude. We have not devoted much attention to this regime -- for example, we have not investigated when damping or growth would occur in these special conditions -- since in the SN decoupling region presumably the opposite limit holds and $\tomegaE$ and  $\Gamma_E$ are small perturbations. 

Our main practical result is an approximate dispersion relation, based on the expansions in Eqs.~\eqref{eq:expansion_subluminal_In} and~\eqref{eq:collisional_subluminal_expansion}, which avoids integrals over energy and direction. It is easily generalized to the mixed slow-collisional regime. Such algebraic dispersion relations are much easier to solve as they avoid multi-dimensional integrals. The validity of this approximation is guaranteed by the condition  $\tomegaE, \Gamma_E \epsilon_\Gamma \ll \mu \epsilon$, which always holds. In this scheme, it is straightforward to determine the corrections due to neutrino masses and collisions in any regime of interest. Moreover, our treatment also illuminates that subluminal modes are the primary cause for fast instabilities, but presumably not of great relevance for slow and collisional instabilities in the absence of an angular crossing.

\section{Near-luminal gapped modes}\label{sec:near_luminal}

If $\omega$ is very close to $k$, the Taylor expansions that we have performed in the previous sections break down, because the denominator peaks around the region $v\simeq 1$ where the neutrino angular distribution changes rapidly. Therefore, the properties of collective modes will be determined here primarily by the neutrino angular distribution along the direction of the flavor waves. We will now introduce the general technique to describe the dispersion relation for these near-luminal modes.

\subsection{Dispersion relation near the light cone}

For a small vacuum energy splitting or collision rate, the $I_n$ integrals diverge logarithmically. More specifically, as $\tomegaE,\Gamma_E\to 0$ and if we write
$\omega=k+\chi$, we can expand in $\chi$ and for near-luminal modes include only the dominant logarithmic terms in the integrals. We first introduced this procedure in Ref.~\cite{Fiorillo:2024bzm} for the fast regime and in Ref.~\cite{Fiorillo:2024pns} for the slow case. As a brief outline, for any integral of the form
\begin{equation}
    I(\psi)=\int \frac{F(v) dv}{1-v+\psi+i\epsilon},
\end{equation}
with $|\psi|\ll 1$, we may extract a regular component
\begin{equation}
    I(\psi)=\int \frac{\left[F(v)-F(1)\right]dv}{1-v+\psi+i\epsilon}+F(1)\int\frac{dv}{1-v+\psi+i\epsilon}.
\end{equation}
In the first integral, we can now safely replace $\psi\to 0$, since the singularity in the denominator is harmless due to the vanishing numerator at $v\to 1$, whereas the second integral, which contains the dominant logarithmic singularity for $\psi \to 0$, can now be performed explicitly. Thus, an immediate generalization of Eq.~(5.1) in Ref.~\cite{Fiorillo:2024pns} yields
\begin{equation}
    I_n=\int dE\,\frac{G(1,E)\log\left(\frac{2k}{\chi-\tomegaE+i\Gamma_E}\right)-\oG(1,E)\log\left(\frac{2k}{\chi+\tomegaE+i\overline{\Gamma}_E}\right)}{k}+\frac{d_n}{k},
\end{equation}
where
\begin{equation}
    d_n= \int_{-1}^{+1} dv\, \frac{D(v) v^n-D(1)}{1-v}
\end{equation}
following the earlier notation.

How to handle the logarithm of their complex arguments depends on the $i\epsilon$ prescription in the denominators of $I_n$. In particular, there is a branch cut when $\mathrm{Im}(\chi)=-\Gamma_E$ (similar for antineutrinos), and the phase prescription is such that
\begin{equation}
    \log\left(\frac{2k}{\delta}\right)\to \log\left(\frac{2k}{|\delta|}\right)-i\pi \;\mathrm{sign}(k)
\end{equation}
when $\delta$ changes sign. This is naturally achieved with the prescription
\begin{equation}
    \log(\delta)=\log|\delta|+i\varphi,
\end{equation}
where $-\pi/2<\varphi<3\pi/2$ is the argument of the complex number $\delta$. Moreover,
we choose $\log k=\log |k|$ for $k>0$ and $\log k=\log |k|+i\pi$ for $k<0$. Notice that the logarithm function has a branch cut for $\mathrm{Im}(\delta)<0$, $\mathrm{Re}(\delta)=0$, as expected. 

With this approximate form of the integrals $I_n$, we can now replace in the full dispersion relation Eq.~\eqref{eq:general_dispersion_relation} to obtain
\begin{eqnarray}\label{eq:dispersion_logarithmic_finite_k}
    \kern-1em&&\int dE\left[G(1,E)\log\left(\frac{2k}{\chi-\tomegaE+i\Gamma_E}\right)-\oG(1,E)\log\left(\frac{2k}{\chi+\tomegaE+i\overline{\Gamma}_E}\right)\right]\bigl(d_0+d_2-2d_1\bigr)
    \nonumber\\[1.5ex]
     \kern-1em&&\kern15em{}+d_0 d_2-d_1^2+\frac{k}{\mu}(d_0-d_2)-\left(\frac{k}{\mu}\right)^2=0.
\end{eqnarray}
Notice that $d_0+d_2-2d_1=D_0-D_1$ in terms of the usual moments $D_n=\int dv D(v) v^n$. For an uncrossed distribution, this quantity is always positive, assuming $D(v)>0$. This form of the dispersion relation generally holds close to the positive half of the light cone, with $\omega=k+\chi$ and $\chi\ll k$. We recall that we focus only on cases without an angular crossing.

We can similarly seek solutions with $\omega=-k+\chi$ (modes moving close to the negative light cone). In this case, we introduce the quantities
\begin{equation}
    e_n=\int_{-1}^{+1}dv\,\frac{D(v) v^n-D(-1) (-1)^n}{1+v},
\end{equation}
so that
\begin{eqnarray}\label{eq:dispersion_logarithmic_finite_k-B}
    &&\int dE\left[G(-1,E)\log\left(\frac{-2k}{\chi-\tomegaE+i\Gamma_E}\right)-\oG(-1,E)\log\left(\frac{-2k}{\chi+\tomegaE+i\overline{\Gamma}_E}\right)\right]\bigl(e_0+e_2+2e_1\bigr)
    \nonumber\\[1.5ex]
    &&\kern15em{}+e_0 e_2-e_1^2-\frac{k}{\mu}(e_0-e_2)-\left(\frac{k}{\mu}\right)^2=0
\end{eqnarray}
becomes the dispersion relation.

Throughout this paper, we focus on axially symmetric distributions and on modes directed along the axis of symmetry. However, the appearance of $G(1,E)$ and $G(-1,E)$ reveals that the results for near-luminal instabilities are much more general. Near-luminal instabilities are entirely resonant in direction, and the resonant neutrinos move exactly along the direction of the flavomon. In contrast, for subluminal instabilities, resonant neutrinos can move within an entire cone defined by $\omega=\bk\cdot \bv$. Therefore, the growth rate of near-luminal instabilities depends primarily on the angular distribution close to the flavomon direction, regardless of its global symmetries.

\subsection{Fast flavor limit}

In the fast flavor limit, and without an angular crossing, the dispersion relation simplifies considerably to the form
\begin{equation}
    D(1)(D_0-D_1)\log\left(\frac{2k}{\chi}\right)+d_0 d_2-d_1^2+\frac{k}{\mu}(d_0-d_2)-\left(\frac{k}{\mu}\right)^2=0.
\end{equation}
Because $D(1)(D_0-D_1)$ is always positive for an uncrossed distribution, there is always a real-valued band for $k\to +\infty$, with $\chi=2k \exp\bigl[{-\frac{k^2}{\mu^2 D(1)(D_0-D_1)}}\bigr]$. One can similarly find the properties of the fast, real-valued modes close to the light cone for $k\to -\infty$. 

\subsection{Slow flavor limit}

In the slow flavor limit, Eq.~\eqref{eq:dispersion_logarithmic_finite_k} becomes
\begin{eqnarray}\label{eq:dispersion_slow_logarithmic}
    &&\int dE\left[G(1,E)\log\left(\frac{2k}{\chi-\tomegaE}\right)-\oG(1,E)\log\left(\frac{2k}{\chi+\tomegaE}\right)\right]\bigl(d_0+d_2-2d_1\bigr)
    \nonumber\\[1.5ex]
    &&\kern15em{}+d_0 d_2-d_1^2+\frac{k}{\mu}(d_0-d_2)-\left(\frac{k}{\mu}\right)^2=0.
\end{eqnarray}
We studied this case previously in the monochromatic limit \cite{Fiorillo:2024pns}, but an extension to general energy distributions is difficult because the quantities $\chi$ and $\tomegaE$ within the logarithms may be comparable. However, we will find that the information gathered for the monochromatic case still allows us to understand the main qualitative properties.

\subsubsection{Broad slow instabilities}

We first seek an unstable solution with $\mathrm{Im}(\chi)\gg \tomegaE$. We are driven here by an analogy with the monochromatic example studied in Ref.~\cite{Fiorillo:2024pns}, where we saw that if the neutrino-antineutrino asymmetry $\epsilon\ll 1$ the growth rate is typically $\mathrm{Im}(\chi)\sim \tomegaE/\epsilon\gg \tomegaE$. 

In discussing the relation with monochromaticity, we should also highlight another point; for a monochromatic spectrum, we previously showed~\cite{Fiorillo:2024pns} that for $k\to+\infty$, there was always a solution with $\chi\to\tomegaE$ and a small growth rate (in the case of inverted ordering for $k>0$, or normal ordering for $k<0$). This solution is understood directly from Eq.~\eqref{eq:dispersion_slow_logarithmic} because the integral over $\delta(E)$ is trivial, and we see that as $k\to \infty$, the divergence of the $k^2$ term in the dispersion relation can be balanced by the divergence of the logarithmic term as $\chi\to \tomegaE$. This argument, however, fails for a general spectrum because the divergence of $\log(\chi-\tomegaE)$ disappears after integration over a sufficiently smooth energy distribution. Therefore, we immediately conclude that a branch of slow and potentially unstable modes, extending up to arbitrarily large $k$, is an artifact of monochromaticity.

Assuming $\chi\gg \tomegaE$ -- as usual our approximations are oriented towards extreme regimes, but they qualitatively capture the behavior also in more intermediate regimes -- we can expand Eq.~\eqref{eq:dispersion_slow_logarithmic} as
\begin{equation}\label{eq:approximate_slow_logarithmic}
    D(1)\log\left(\frac{2k}{\chi}\right)+\frac{S(1)}{\chi}+\frac{\mu^2 (d_0 d_2-d_1^2)+\mu k(d_0-d_2)-k^2}{\mu^2(d_0+d_2-2d_1)}=0.
\end{equation}
This equation exactly parallels Eq.~(5.17) in Ref.~\cite{Fiorillo:2024pns} for the monochromatic case. In the regions where the growth rate is the largest, we expect $\chi\sim i \chi_I$, dominated by its imaginary part; in this case, taking the imaginary part of Eq.~\eqref{eq:approximate_slow_logarithmic}, we find
\begin{equation}\label{eq:approximate_growth_rate_slow}
    \chi_I=-\frac{2S(1)}{\pi D(1)}.
\end{equation}
Inverted hierarchy means $S(1)<0$, so we find an instability, and the typical growth rate is given by a formula identical to Eq.~(5.19) in Ref.~\cite{Fiorillo:2024pns} for the monochromatic case, except that the vacuum frequency is averaged over the spectrum, as implicit in the definition of $S(1)$. For $k<0$, the sign of $\chi_I$ flips, and therefore an instability appears in normal, rather than inverted, ordering.

In fact, this reasoning can be made somewhat more general by re-arranging the terms in the approximate dispersion relation as $\phi_\chi(\chi)=\phi_k(k)$,
where
\begin{subequations}
\begin{eqnarray}
     \phi_\chi(\chi)&=&D(1)\log\chi-\frac{S(1)}{\chi},
     \\
     \phi_k(k)&=&D(1)\log(2k)+\frac{\mu^2 (d_0 d_2-d_1^2)+\mu k(d_0-d_2)-k^2}{\mu^2(d_0+d_2-2d_1)}.
\end{eqnarray}    
\end{subequations}
For $k>0$, $\phi_k(k)$ is certain to be real, and therefore a solution for complex $\chi$ exists only if $\mathrm{Im}[\phi_\chi(\chi)]=0$. For $\chi=|\chi| e^{i\varphi}$,
where  $-\pi/2<\varphi<3\pi/2$ from the definition of the logarithm, we find that 
\begin{equation}
    \frac{\sin \varphi}{\varphi}=-\frac{D(1) |\chi|}{S(1)}.
\end{equation}
Thus, for $S(1)>0$ (normal ordering), there are solutions with $\pi<\varphi<3\pi/2$, which therefore are damped; while for $S(1)<0$ (inverted ordering), there are also growing ones. For $k<0$, the situation is inverted. The approximate solutions can be given as
\begin{equation}
    \chi(\varphi)=-\frac{S(1) \sin\varphi}{D(1)\varphi}e^{i\varphi}
\end{equation}
in parametric form.

Finally, we notice that for a continuous energy distribution, the dispersion relation in Eq.~\eqref{eq:dispersion_slow_logarithmic}, strictly speaking, never admits a purely real-valued solution, in direct contrast with the purely real branch we had found in the monochromatic case~\cite{Fiorillo:2024pns}. In Eq.~\eqref{eq:dispersion_slow_logarithmic}, if $\chi$ is real, in one part of the integration range one has $\chi-\tomegaE>0$, whereas in another part one has $\chi-\tomegaE<0$. For negative $\chi$, the same issue arises for $\chi+\tomegaE$ in the other logarithmic term. For negative argument, the logarithm acquires an imaginary part $-i\pi$, so that the dispersion relation unavoidably becomes complex. 

In the monochromatic case, this issue does not arise because there is a single frequency and therefore the real-valued branch of Ref.~\cite{Fiorillo:2024pns} had $\chi>\tomegaE$ always. We now discover that gapped modes always have a small imaginary part, due to their resonance with a small fraction of neutrinos with a sufficiently large $\tomegaE$. Of course, since the number of neutrinos at small $E$, and therefore at large $\tomegaE$, decreases as a power law of $E$, this imaginary part is correspondingly suppressed. Therefore, in the non-monochromatic case, even the branch without instability is expected to show Landau damping.

The expression we have just found applies when $\epsilon\ll 1$, but we can give it more generality. For it to apply, we require $\chi_I\gg \tomegaE$; if neutrinos and antineutrinos are very symmetrical, with $\epsilon \ll 1$, then $D(1)$ is very small, ensuring this inequality holds, but this can also be achieved with different setups. The physical meaning of the inequality is that the width of the flavomon energy level, measured by $\chi_I$, is much larger than the typical neutrino energy splitting caused by their masses, so that the instability is caused by an interaction between flavomons and neutrinos non-resonant in energy. Flavomons interact with the entirety of the neutrino energy distribution at once, which is why the growth rate depends essentially only on energy-integrated quantities. We emphasize the meaning of resonant here as connected to the width of flavomon-neutrino interaction, not to a particular enhancement of the growth rate. The nomenclature of resonance somewhat fails us here because it can be used with many different meanings. Furthermore, we stress that, while the flavomon-neutrino interaction is non-resonant in energy (the width of the level is large compared with the neutrino energy splitting), it is instead resonant in direction. The energy splitting between neutrinos with different directions is of the order of $kv\sim \mu \epsilon\gg \chi_I$, which is why we can associate a specific growth rate with each neutrino direction; its physical meaning is the growth rate of flavomons collinear with those neutrinos.

\subsubsection{Narrow slow instabilities}

We now turn to the opposite regime in which the growth rate is so small that the flavomon-neutrino interaction is resonant also in energy. This corresponds to a \textit{weak} slow instability, so that the neutrino distribution by a slight deformation could be turned into a stable one. We will now find the conditions for this to occur, always assuming the regime $\tomegaE\ll \mu \epsilon$ that allows us to treat the vacuum frequency perturbatively and to use Eq.~\eqref{eq:dispersion_slow_logarithmic}. It can be rearranged in the form
\begin{equation}
    \int dE \left[ G(1,E) \log\left(\frac{2k}{\chi-\tomegaE}\right)-\oG(1,E) \log\left(\frac{2k}{\chi+\tomegaE}\right) \right]=f(k),
\end{equation}
where the specific form of the function $f(k)$ is inessential.

To be specific, we focus on normal ordering ($\tomegaE>0$) and $k<0$; it is straightforward to generalize to other cases. A slow-unstable mode corresponding to a narrow resonance must have $\chi_I=\mathrm{Im}(\chi)\ll \chi_R=\mathrm{Re}(\chi)$. Therefore, we can proceed to expand the logarithms in a series of $\chi_I$. Taking the imaginary part of the dispersion relation we then find
\begin{eqnarray}
    \kern-4em&&\mathrm{Im}\left\{\int dE \left[ G(1,E) \log\left(\frac{2k}{\chi_R-\tomegaE}\right)-\oG(1,E) \log\left(\frac{2k}{\chi_R+\tomegaE}\right) \right]\right\}={}
    \nonumber\\ 
    \kern-4em&&\kern18em{}=\fint  dE\left[\frac{G(1,E)}{\chi_R-\tomegaE}-\frac{\oG(1,E)}{\chi_R+\tomegaE}\right]\chi_I.
\end{eqnarray}
The imaginary part of the integrals on the left-hand side comes entirely from the region of integration where the argument of the logarithm is negative. Assuming $\chi_R>0$, finally yields
\begin{equation}\label{eq:approximate_growth_rate_slow_resonant}
    \chi_I=\pi\frac{\int dE[G(1,E) \Theta(\chi_R-\tomegaE)-\oG(1,E)]}{\fint  dE\left[\frac{G(1,E)}{\chi_R-\tomegaE}-\frac{\oG(1,E)}{\chi_R+\tomegaE}\right]},
\end{equation}
where $\Theta(x)$ is the Heaviside function. Instead, for $\chi_R<0$, the neutrino part of the integral is purely real, and therefore we have
\begin{equation}
    \chi_I=-\pi\,\frac{\int dE \oG(1,E) \Theta(\chi_R+\tomegaE)}{\fint  dE\left[\frac{G(1,E)}{\chi_R-\tomegaE}-\frac{\oG(1,E)}{\chi_R+\tomegaE}\right]}.
\end{equation}
These expressions provide a direct approximation for the growth rate of slow-unstable modes in the narrow approximation, i.e., when the flavomon-neutrino interaction is resonant both in energy and direction. In this sense, they can be directly compared in spirit to the resonant approximation for the growth rate of fast-unstable modes that we derived in Ref.~\cite{Fiorillo:2024uki}. Interestingly, in the slow case, the growth rate depends not on the energy distribution itself, but rather on its integral conditioned by the Heaviside function caused by the logarithmic term in the approximate dispersion relation. 

For a distribution without energy crossing, the numerator of Eq.~\eqref{eq:approximate_growth_rate_slow_resonant} has a fixed sign and can never vanish. Hence, in its regime of validity, the collective modes for varying $\chi_R$ can never transition from damped to unstable or vice versa. We can also see directly that in fact they are always damped; for $\chi_R\to 0$ a non-crossed distribution has necessarily the numerator and the denominator in Eq.~\eqref{eq:approximate_growth_rate_slow_resonant} with opposite signs, and therefore $\chi_I<0$. Therefore, there can only be weakly damped, but not weakly unstable, modes for an uncrossed distribution. Ultimately, a global argument based on lepton number conservation~\cite{Fiorillo:2024bzm, Fiorillo:2024uki} shows directly that no instability can ensue, either weak or strong; this local argument, while more restricted to the case of weak instability only, provides a kinematic perspective on this conclusion.

If an energy crossing exists, then a value of $\chi_R$ for which the numerator of Eq.~\eqref{eq:approximate_growth_rate_slow_resonant} can exist. At such value of $\chi_R$, there will be a purely stable mode, whereas infinitesimally to the right or left, $\chi_I$ will be either positive or negative, so that on one side of this value unstable modes are certain. This is precisely the same dynamics that leads to the formation of fast-unstable modes. We will show in Sec.~\ref{sec:narrow_slow_numerical} that indeed Eq.~\eqref{eq:approximate_growth_rate_slow_resonant} provides an excellent approximation for the growth rate of unstable modes in the narrow-resonance regime.

\subsection{Collisional flavor limit}

The collisional limit for modes close to the light cone has never been investigated. The dispersion relation Eq.~\eqref{eq:dispersion_logarithmic_finite_k} becomes now
\begin{eqnarray}\label{eq:dispersion_logarithmic_collisional}
    \kern-4em&&\int dE\left[G(1,E)\log\left(\frac{2k}{\chi+i\Gamma_E}\right)-\oG(1,E)\log\left(\frac{2k}{\chi+i\overline{\Gamma}_E}\right)\right]\bigl(d_0+d_2-2d_1\bigr)
    \nonumber\\[1.5ex]
    \kern-4em&&\kern17em{}+d_0 d_2-d_1^2+\frac{k}{\mu}(d_0-d_2)-\left(\frac{k}{\mu}\right)^2=0.
\end{eqnarray}
Once again, in the non-monochromatic case, we do not expect any solution in the limit $k\to \infty$; even if $\chi$ were chosen to make the argument of the logarithm $\log(\chi+i\Gamma_E)$ vanish for a single energy, after integrating over a smooth spectrum, the divergence disappears. Therefore, just as in the slow case, we focus on seeking solutions with $\chi\gg\Gamma$ as a starting point. Our rationale here follows the analogy with the superluminal and subluminal modes; while the corrections to the eigenfrequencies due to neutrino masses are always of the order of $\tomegaE/\epsilon$, the analogous corrections due to collisions are of the order of $\Gamma_E \epsilon_\Gamma/\epsilon$. Therefore, again we may take $\epsilon\lesssim1$ as a starting point for a qualitative approximation. By performing this expansion, we find the dispersion relation
\begin{equation}
    D(1)\log\left(\frac{2k}{\chi}\right)-\frac{i\Sigma(1)}{\chi}+\frac{\mu^2 (d_0 d_2-d_1^2)+\mu k(d_0-d_2)-k^2}{\mu^2(d_0+d_2-2d_1)}=0,
\end{equation}
similar to the slow case.

Again we re-arrange the terms in this dispersion relation in the form $\phi_\chi(\chi)=\phi_k(k)$, but now find
\begin{equation}
    \phi_\chi(\chi)=D(1)\log\chi+\frac{i\Sigma(1)}{\chi},
\end{equation}
so that with $\chi=|\chi|e^{i\varphi}$, we find
\begin{equation}
    \frac{\cos\varphi}{\varphi}=-\frac{D(1)|\chi|}{\Sigma(1)}.
\end{equation}
Thus, if $D(1)/\Sigma(1)<0$, we have $0<\varphi<\pi/2$, while if $D(1)/\Sigma(1)>0$ we have $-\pi/2<\varphi<0$ or $\pi/2<\varphi<3\pi/2$. In both cases, there can also be growing solutions, so the sign of $\Sigma(1)$ does not discriminate whether an unstable collisional mode exists or not based, whereas for slow modes it did. We will later see in explicit examples that, since the near-luminal modes must smoothly connect with the superluminal ones, near-luminal unstable collisional modes can exist if there are superluminal unstable ones.

\subsection{Summary}

In the fast limit, near-luminal modes can become unstable if an angular crossing is present. Such a distribution is dominated by these fast instabilities, which are therefore discussed in greater detail in the theory of fast flavor evolution~\cite{Fiorillo:2024bzm, Fiorillo:2024uki, Fiorillo:2024dik}. In the absence of angular crossings, near-luminal modes remain stable in the fast limit.

In the slow limit, near-luminal modes are the primary class of collective oscillations that become unstable. This is intuitively understandable: subluminal modes are strongly Landau-damped, so the small vacuum frequency is insufficient to destabilize them, while superluminal modes are not resonant with any particles and therefore cannot grow -- individual (anti)neutrinos cannot emit flavomons on-shell. Near-luminal modes are stable in the fast limit; however, when the neutrino dispersion relation is modified by the vacuum frequency, the corresponding flavomons can be emitted on-shell. 

The growth rate of the instability coincides with the width of the flavomon energy level; if the width far exceeds the energy splitting (broad slow instability), the growth rate is enhanced by the flavomon interaction with the whole energy distribution, so we might call this an instability non-resonant in energy. In the opposite case of narrow slow instability, the energy distribution is close to stability, and the growth rate is determined by the resonant interaction with specific energies associated with a crossing.

In the collisional limit, near-luminal modes exist and can be derived from the approximate dispersion relation in Eq.~\eqref{eq:dispersion_logarithmic_collisional}. These near-luminal modes must connect with the superluminal ones discussed in Sec.~\ref{sec:superluminal_collisional}. Therefore, if the superluminal modes are unstable, this is also expected for the near-luminal ones. We will confirm this in Sec.~\ref{sec:example_distributions}, where explicit examples of the collective eigenmodes for specific neutrino distributions will be presented.

\section{Gapless modes}\label{sec:gapless}

We have so far assumed that $\mathrm{Re}(\omega)\gg \tomegaE, \Gamma_E$. Each such mode arises from a definite solution of the dispersion relation in the fast limit, which has $\mathrm{Re}(\omega)\sim \mu \epsilon\gg \tomegaE, \Gamma_E$. In other words, these slow or collisional modes arise from deformations of corresponding fast ones, to which they are connected in a continuous manner. However, it is also possible that new slow and collisional solutions emerge that did not exist in the fast limit and for them we expect $\mathrm{Re}(\omega)\sim \tomegaE,\, \Gamma_E$. For $\tomegaE,\Gamma_E\to 0$, their eigenfrequencies formally vanish -- hence our choice of naming them gapless -- and they physically disappear. For such gapless modes, a distinction between subluminal, near-luminal, and superluminal waves is less helpful, since a mode may be subluminal with respect to neutrinos with certain energies, and superluminal with respect to others. 

\subsection{Dispersion relation}

From the definition of gapless modes as having $\mathrm{Re}(\omega)\sim \tomegaE,\, \Gamma_E\ll \mu \epsilon$, we can directly deduce several of their properties. Since now there is no term in the denominators of the integrals $I_n$ which is of order $\mu \epsilon$, in the complete dispersion relation Eq.~\eqref{eq:general_dispersion_relation} all terms of the form $\mu I_n\sim \mu/\tomegaE \gg 1$ in the slow limit, or $\mu I_n\sim \mu/\Gamma_E\gg 1$ in the collisional one. Therefore, all terms of order 1 in the dispersion relation can be neglected in first approximation, leading to the homogeneous polynomial
\begin{equation}\label{eq:dispersion_relation_slow_only}
    I_0 I_2=I_1^2.
\end{equation}
Notice that $\mu$ has dropped out of the dispersion relation altogether, as expected since the eigenmodes we are seeking do not exist in the fast limit, and therefore their wavenumber and frequency are independent of $\mu$. 

Gapless eigenmodes have appeared scattered over the previous literature: in particular, a collisional unstable mode with $\omega\ll\mu\epsilon$ is explicit in Refs.~\cite{Xiong:2022zqz, Lin:2022dek, Fiorillo:2023ajs}. Moreover, in the theory of slow conversions~\cite{Fiorillo:2024pns, Fiorillo:2025ank} we showed the emergence of real-valued modes with very small frequency in the limit of $\tomegaE\ll \mu \epsilon^2$ (see, e.g., the zoomed-in panels in Figs.~2 and~3 of Ref.~\cite{Fiorillo:2024pns}). These different cases seemed completely unrelated, whereas our unified perspective shows that they actually belong to the same class of gapless modes, with properties that can be understood from the dispersion relation. 

From the analytical perspective, gapless modes are much more complicated because there is no hierarchy of scales. Both the eigenfrequency and wavenumber are of the same order of magnitude as $\tomegaE$ and/or $\Gamma_E$, which also has physical consequences. For gapped modes, since $\mu\epsilon$ was the largest scale, neutrinos resonant with flavomons were determined primarily by their direction. In the resonance condition $\omega-\bk\cdot\bv-\tomegaE=0$, vacuum frequencies were a small perturbation, negligible except when $\omega\simeq |\bk|$, which is why slow instabilities appear primarily close to the light cone. However, if $\omega\sim |\bk|\sim \tomegaE$, there is no well-defined notion of a light cone, since neutrinos with different energies have different ``light cones'' and the resonance depends intrinsically on both neutrino energy and direction. As a consequence, general results for gapless modes are hard to obtain. 

Therefore, we restrict our analytical arguments to $k=0$ and isotropic distributions. The dispersion relation then factorizes in two separate conditions, $I_0=0$ and $I_2=0$, which are actually identical for an isotropic distribution. In other words, gapless monopole and quadrupole modes have approximately the same dispersion given by $I_0=0$, or explicitly
\begin{equation}
    \int  \frac{G_0(E)\,dE}{\omega-\tomegaE+i\Gamma_E}-\int \frac{\oG_0(E)\,dE}{\omega+\tomegaE+i\overline{\Gamma}_E}=0.
\end{equation}
While this condition is based on very restrictive assumptions, the results are qualitatively more general as suggested by our numerical examples in Sec.~\ref{sec:example_distributions}.

The apparent simplicity of this dispersion relation is actually deceptive, but there is at least one regime that allows for an analytic understanding: the near-symmetric case $\epsilon\ll 1$. While this may not be a particularly realistic condition, it offers some qualitative insight, and therefore is our initial case.

\subsection[Near-symmetric regime: \texorpdfstring{$\epsilon\ll 1$}{}]{Near-symmetric regime: \texorpdfstring{\boldmath$\epsilon\ll 1$}{}}
\label{sec:gapless_near_symmetric}

If the neutrino and antineutrino densities are very similar ($\epsilon\ll1$), we have repeatedly found that the typical eigenfrequencies in the slow or collisional cases are not $\tomegaE$ or $\Gamma_E$, but enhanced by a factor $\epsilon^{-1}$, suggesting $\omega \gg \tomegaE, \Gamma_E$ as a starting assumption. Moreover, as explained earlier, we consider only $k=0$ and an isotropic distribution, implying $I_1=0$. It is straightforward to generalize our expansion, but this exercise would not be particularly illuminating because we aim only at a qualitative understanding of the gapless modes.

For $I_1=0$, the dispersion relation factorizes as usual into monopole ($I_0=0$) and quadrupole modes ($I_2=0$). Focusing on the former and expanding for $\omega\gg \tomegaE, \Gamma_E$, we find the dispersion relation to be
\begin{equation}\label{eq:gapless_dispersion_relation}
    \frac{D_0}{\omega}+\frac{S_0-i\Sigma_0}{\omega^2}=0.
\end{equation}
The difference to gapped superluminal modes is a vanishing right-hand side, as behooves a gapless solution. One immediately finds the simple expression
\begin{equation}\label{eq:gapless_near_symmetric}
    \omega=\frac{i\Sigma_0-S_0}{D_0},
\end{equation}
which reflects most of the qualitative features of gapless modes. 

In the slow limit ($\Sigma_0\to 0$), the frequency is purely real and is of order $\omega\sim \tomegaE/\epsilon$ as expected, confirming the self-consistency of our expansion for $\epsilon\ll1$. While the integrals $I_n$ acquire a small imaginary part from the energy range around $\omega=\tomegaE$, by the $\epsilon\ll1$ assumption, this part of the integration range will contain very few neutrinos and can be neglected. Consequently, for $\epsilon\ll 1$, slow gapless modes are essentially real valued.

In the collisional limit ($S_0\to 0$), the eigenfrequency is purely imaginary; the mode is damped for $\Sigma_0<0$ and unstable for $\Sigma_0>0$. This condition is opposite to superluminal gapped modes, which instead are unstable for $\Sigma_0>0$. For a realistic supernova core, we expect $\Sigma_0>0$, since there are more $\nu_e$ than $\overline\nu_e$, and the former interact more strongly. Therefore, a collisional instability in this environment is likely to be of the gapless kind. 

\subsection[General case: \texorpdfstring{$\epsilon \sim 1$}{}]{General case: \texorpdfstring{\boldmath$\epsilon \sim 1$}{}}

For $\epsilon\sim 1$, these approximate solutions no longer apply, but they remain useful as a starting point for qualitative understanding: we can imagine to start with a distribution with $\epsilon \ll 1$, and slowly increase $\epsilon$. We proceed to discuss how the solutions change as $\epsilon$ increases, studying separately
the slow and collisional limits.

\subsubsection{Slow modes}

For slow modes, as $\epsilon$ grows, $\mathrm{Re}(\omega)$ becomes comparable with a typical $\tomegaE$ within the energy distribution. Therefore, the integrals $I_n$ acquire sizable imaginary parts, and in turn, the gapless slow modes, which previously were real-valued, also acquire an imaginary part. We can prove that it is negative (the modes are damped) using a version of Nyquist's argument applied to the energy integral, rather than the velocity integral as we had previously done for the fast instability \cite{Fiorillo:2023hlk, Fiorillo:2024dik}. In the present case, the dispersion relation has the form
\begin{equation}
    \Phi(\omega)=I_0(\omega) =0.
\end{equation}
It is straightforward to extend the argument to a non-isotropic distribution and use the full dispersion relation of Eq.~\eqref{eq:dispersion_relation_slow_only}, but for simplicity we stick to isotropy.

According to Nyquist's argument, reviewed in more detail in Refs.~\cite{Fiorillo:2023hlk,Fiorillo:2024dik}, for the function $\Phi(\omega)$ to have a zero in the upper half-plane of the complex variable $\omega$, its trajectory in the complex plane of $\Phi$ as $\omega$ runs from $-\infty$ to $+\infty$ must wrap at least once around the origin. Compared with the velocity integral~\cite{Fiorillo:2023hlk, Fiorillo:2024dik}, the present case is even simpler, since the variable $\omega$ is not restricted in a finite interval, and therefore the integrand function has no discontinuities. For real $\omega$, the function $\Phi(\omega)$ is generally complex, since
\begin{equation}
    I_0=\fint \frac{G_0(E) dE}{\omega-\tomegaE}-i\pi\int G_0(E) \delta(\omega-\tomegaE) dE-\fint \frac{\oG_0(E) dE}{\omega+\tomegaE}+i\pi\int G_0(E) \delta(\omega+\tomegaE) dE,
\end{equation}
following the usual prescription from causality. If $G_0(E)$ and $\oG_0(E)$ never vanish, as expected if there are much fewer $\nu_\mu$ and $\overline\nu_\mu$ than their electronic partners, then the imaginary part of $I_0$ never vanishes, except at $\omega=0$, where it changes sign. At any other frequency, the integrals always have non-vanishing imaginary parts, even if very small. Consequently, the trajectory of $\Phi$ in the complex plane starts for $\omega\to -\infty$ at $\Phi(\omega)\to 0$, and then can cross the real axis only once before returning to $0$ at $\omega\to +\infty$. Therefore, it can never wrap fully around the origin. It was crucial here to use the homogeneous nature of the dispersion relation, such that $\Phi(\omega)\to 0$ for $\omega\to \pm\infty$. If the dispersion relation were inhomogeneous, the function could start at a nonzero value on the real axis, and therefore could encircle the origin even with a single crossing. Indeed, the inhomogeneous dispersion relation can admit unstable solutions for $\epsilon\ll 1$, which are the superluminal gapped solutions with $\mathrm{Im}(\omega)\sim \sqrt{\mu \tomegaE}$ that we have examined in Sec.~\ref{sec:superluminal}.

Therefore, no slow unstable gapless modes exist for distributions with both $G_0(E)>0$ and $\oG_0(E)>0$ and indeed, in our numerical examples in Sec.~\ref{sec:example_distributions}, no such modes turn up. They might be supported by more intricate energy distributions, but we do not further study this question that seems phenomenologically less motivated. 

\subsubsection{Collisional modes}

We have seen that in the physical case of neutrinos colliding more frequently than antineutrinos, in the limit $\epsilon\ll 1$, gapless modes are collisionally unstable, but with increasing $\epsilon$, these instabilities must eventually disappear. This behavior is most easily understood in a simple model of a monochromatic energy distribution, as recognized earlier~\cite{Johns:2021qby}. Using $\mathcal{G}=\int dE\, G_0(E)$ and $\overline{\mathcal{G}}=\int dE \, \oG_0(E)$, the dispersion relation simplifies to
\begin{equation}
    \frac{\mathcal{G}}{\omega+i\Gamma}-\frac{\overline{\mathcal{G}}}{\omega+i\overline{\Gamma}}=0,
\end{equation}
where $\Gamma$ and $\overline{\Gamma}$ are the damping rate $\Gamma_E$ and $\overline{\Gamma}_E$ evaluated at the fixed (anti)neutrino energy. Solving this equation directly reveals that it admits a growing mode only if
\begin{equation}
    \epsilon=\frac{\mathcal{G}-\overline{\mathcal{G}}}{\mathcal{G}+\overline{\mathcal{G}}}<\frac{\Gamma-\overline{\Gamma}}{\Gamma+\overline{\Gamma}}.
\end{equation}
For non-monochromatic modes, we then also expect gapless collisional instabilities to disappear when $\epsilon$ becomes sufficiently large.

To develop a precise criterion, we imagine to distort the antineutrino distribution as $\oG_0(E)\to f \oG_0(E)$, with the factor $f$ chosen such that for $f=1$ the number density of neutrinos and antineutrinos are the same ($\epsilon \sim 0$). In this low-$\epsilon$ limit, an unstable mode is certain to exist (Sec.~\ref{sec:gapless_near_symmetric}), and it has a purely imaginary eigenfrequency. Next we imagine to lower $f$, thereby increasing $\epsilon$, and slowly distorting
the unstable mode. It can disappear only when the eigenfrequency passes through 0, the point where the analytical properties of the dispersion relation change. Therefore, this critical value of $f$ must be determined from the condition
\begin{equation}\label{eq:condition_disappearance_collisional}
    \int \frac{G_0(E) dE}{\Gamma_E}-\int \frac{f\oG_0(E) dE}{\overline{\Gamma}_E}=0.
\end{equation}
For lower values of $f$ (larger values of $\epsilon$), there will be no unstable collisional eigenmode of the gapless kind.

\subsection{Summary}

Gapless modes exist only in the slow and collisional limits. For the first time, we have provided an approximate dispersion relation (Eq.~\ref{eq:dispersion_relation_slow_only}) that, by definition, describes only gapless modes, as $\mu$ does not appear in it at all.

In the slow limit, these modes are generally stable or subject to Landau damping. Flavor waves resonate with neutrinos of specific energies (for $k=0$), which can therefore absorb them. In the near-symmetric limit ($\epsilon\to 0$), the flavomon energy is $\omega\sim \tomegaE/\epsilon\gg \tomegaE$, and thus the fraction of neutrinos with a sufficiently large vacuum frequency to meet the resonance condition is very small. These modes are essentially stable. As $\epsilon$ increases, the flavomon frequency decreases, allowing them to interact resonantly with neutrinos and become Landau-damped. In astrophysical contexts, these modes are not particularly interesting, as they cannot grow.

In the collisional limit, on the other hand, gapless modes are potentially the most interesting. When neutrinos scatter more frequently than antineutrinos, these are, in fact, the only modes that can become unstable. In the near-symmetric limit ($\epsilon\to0$), their growth rate is given by Eq.~\eqref{eq:gapless_near_symmetric}, which provides an approximate estimate for the scale of collisional instabilities and agrees with the earlier literature~\cite{Xiong:2022zqz}. As the asymmetry increases, this approximation becomes ever less accurate, until the collisional mode disappears entirely.  In Ref.~\cite{Xiong:2022zqz}, this condition was approximately found for a monochromatic neutrino gas as a benchmark, whereas Eq.~\eqref{eq:condition_disappearance_collisional} provides the exact condition for a general energy distribution. Thus, if the neutrino-antineutrino density is not sufficiently symmetric, collisionally unstable modes may disappear altogether. 

It would be straightforward to generalize these results to non-isotropic distributions. The condition for the disappearance of collisional instability -- namely, the existence of a mode with zero eigenfrequency -- still holds. In practice, collisionally unstable modes are most relevant in regions where $\Gamma_E\gtrsim \tomegaE$, where anisotropies are likely small. Therefore, we do not examine these questions in greater detail.

\section{General summary and example distributions}\label{sec:example_distributions}

In this section, we provide an overview and summary of the analytic findings of the previous sections and explicitly demonstrate the main anticipated features for concrete examples.

\subsection{Schematic overview}

To justify our choice of examples, we first summarize our findings. Table~\ref{tab:modes} provides an overview of the different classes of modes we have identified and their behavior in various limits. The table is intended only as a schematic summary, highlighting which modes may become unstable. We do not discuss in detail the existence of additional stable or damped modes. For instance, in the slow limit, alongside the unstable near-luminal gapped branch, a Landau-damped branch also exists, as we have previously discussed in Refs.~\cite{Fiorillo:2024pns, Fiorillo:2025ank} and as will be further illustrated below.

\begin{table}[ht]
\caption{Nature of modes according to their classification. We assume $\tomegaE,\Gamma_E\ll \mu \epsilon^2$, otherwise the near-zero lepton number enhances the growth rate as discussed in Sec.~\ref{sec:superluminal}. We also assume the absence of an angular crossing, otherwise subluminal and near-luminal modes close to the ``flipped'' region of the crossing are unstable.
\label{tab:modes}}
\vskip6pt
\setlength\tabcolsep{0pt}
\begin{tabular*}{\linewidth}{@{\extracolsep{\fill}}llll}
\toprule
& \textbf{Fast} & \textbf{Slow} & \textbf{Collisional} \\
\midrule
\textbf{Gapped modes}&\\
\quad\textbf{-- Superluminal}&Stable&Stable&Damped (for $\Gamma>\overline{\Gamma}$)\\
&&&Unstable (for $\Gamma<\overline{\Gamma}$)\\
\quad\textbf{-- Subluminal}  &Landau-damped&Landau-damped&Landau-damped\\
\quad\textbf{-- Near-luminal}&Stable&Unstable&Damped or unstable\\
\textbf{Gapless modes}       &Non-existent&Landau-damped&Damped (for $\Gamma>\overline{\Gamma}$)\\
&(by definition)&&Unstable (for $\Gamma<\overline{\Gamma}$)\\
\bottomrule
\end{tabular*}
\end{table}

Unstable modes appear at large wavenumbers $k$ and near the light cone when triggered by the vacuum frequency, whereas they emerge already at $k = 0$ when triggered by collisions. When neutrinos scatter more rapidly than antineutrinos, as expected in a SN core, collisionally unstable modes begin at $k = 0$ as gapless modes. In the opposite case, they begin as gapped modes with a large $\text{Re}(\omega)$. In both scenarios, these modes can, in principle, continue as unstable modes near the light cone. Thus, near-luminal modes in the collisional limit might be either damped or unstable. As we will demonstrate through our examples, they generally tend toward damping at large $k$.

The details of this general behavior depend in turn on the parameter $\epsilon$, roughly measuring the asymmetry between neutrinos and antineutrinos:
\begin{itemize}
\item For $\epsilon\ll \sqrt{\tomegaE/\mu},\sqrt{\Gamma_E \epsilon_\Gamma/\mu}$, the neutrino-antineutrino distribution is highly symmetric, implying that unstable modes appear in the slow regime with a growth rate $\mathrm{Im}(\omega)\sim \sqrt{\tomegaE \mu}$, the slow flavor pendulum being the prototype example. Instabilities also arise in the collisional regime, with a growth rate $\mathrm{Im}(\omega)\sim \sqrt{\Gamma_E \epsilon_\Gamma \mu}$, sometimes called resonance-like collisional instabilities. We mostly ignore such small values of $\epsilon$, typically below a few percent, although we cannot conclusively rule out this range in astrophysical environments.

\item Larger $\epsilon$, but still $\epsilon \ll 1$, represents the near-symmetric case, in which the corrections induced by $\tomegaE$ or $\Gamma_E$ are of the order of $\tomegaE/\epsilon$ or $\Gamma_E/\epsilon$, meaning that they are enhanced by the near-symmetry of the distribution. The analytical approximations of the previous sections are here particularly well-suited even for gapless modes, because $\epsilon$ provides a small parameter in the problem.

\item Finally, for $\epsilon \sim 1$, corrections to $\omega$ are of  order $\tomegaE$ or $\Gamma_E$. Analytical approximations for gapless modes become generally unreliable, while they still hold for gapped modes since $\mathrm{Re}(\omega)$ remains very large compared to all other scales. On the other hand, in the slow limit, gapless modes are anyway stable or Landau-damped, diminishing their phenomenological relevance. In the collisional limit, the collisionally unstable mode disappears completely when a critical value for $\epsilon$ is reached. We have explicitly identified this point by the condition of a zero-eigenfrequency mode.

\end{itemize}

\subsection{Reference distributions}

After this summary, we now exemplify our general results by numerical investigations of the instability patterns for two reference distributions. The spectral shapes are taken to follow Maxwell-Boltzmann distributions with different temperatures $T$ and $\overline T$ for neutrinos and antineutrinos, but the exact spectral form is not important. Neutrinos are taken isotropic, whereas antineutrinos are taken to show a small anisotropy measured by a parameter $\epsilon_v$. Specifically, we propose
\begin{equation}\label{eq:distributions}
 G(E,v)=\frac{1}{2}\,\mathcal{N} \left(\frac{E}{T}\right)^2 \frac{e^{-E/T}}{2T}
 \quad\mathrm{and}\quad
 \oG(E,v)=\frac{1+\epsilon_v v}{2}\,\overline{\mathcal{N}} \left(\frac{E}{\overline{T}}\right)^2 \frac{e^{-E/\overline{T}}}{2\overline{T}},
\end{equation}
where the normalization parameters
\begin{equation}
\mathcal{N}= \frac{n_\nu}{n_\nu+n_{\overline{\nu}}}=\int dE dv G(E,v)
\quad\mathrm{and}\quad
\overline{\mathcal{N}}=\frac{n_{\overline{\nu}}}{n_\nu+n_{\overline{\nu}}}=\int dE dv \oG(E,v),
\end{equation}
measure the relative neutrino and antineutrino number density. 

We set $\mu=1$ as a reference energy scale. Moreover, we use a parameterized form for the vacuum frequency and the collision rates of the form
 \begin{equation}
 \tomegaE=\frac{\omega_0}{E/1\; \mathrm{MeV}},
 \quad 
 \Gamma_E=\Gamma_0 \left(\frac{E}{1\;\mathrm{MeV}}\right)^2,
 \quad
 \overline{\Gamma}_E=\overline{\Gamma}_0  \left(\frac{E}{1\;\mathrm{MeV}}\right)^2.
 \end{equation}
The quadratic energy dependence of the collision rates is meant to mimic the approximate behavior of beta interactions with nucleons, but in any case, the qualitative picture depends only mildly on these details. All of these terms are understood in units of $\mu$ and thus dimensionless after setting $\mu=1$, whereas the unit MeV is only notional to show the energy variation of these expressions.

The values of $\omega_0$, $\Gamma_0$, and $\overline{\Gamma}_0$ must, of course, be chosen to respect the hierarchies discussed in Sec.~\ref{sec:scales}. The specific values for our two representative cases are:
\begin{itemize}
\item \textbf{Near-symmetric case}---representative for $\epsilon \ll 1$. Here we assume perfect isotropy for both components ($\epsilon_v=0$), $T=\overline{T}=1\;\mathrm{MeV}$, and $n_{\overline{\nu}}=0.99\,n_{{\nu}}$. Since the angular distribution is isotropic, there is no angular crossing, so we exclude fast instabilities.

\item \textbf{Asymmetric case}---in which $\epsilon \sim 1$. Specifically, $\epsilon_v=0.1$, representative of antineutrinos being slightly more anisotropic than neutrinos in the SN decoupling regions, and different $T=1\; \mathrm{MeV}$ and $\overline{T}=2\;\mathrm{MeV}$, motivated by antineutrinos having higher energies than neutrinos in a SN, and finally $n_{\overline{\nu}}=0.1\,n_\nu$. Once more,
there is no angular crossing and no fast instability.
\end{itemize}

For both of these cases, we consider the slow limit ($\Gamma_E=\overline{\Gamma}_E=0$) and the collisional one ($\tomegaE=0$). In the former, we distinguish between normal mass ordering (NO) with $\tomegaE>0$ and inverted ordering (IO) with $\tomegaE<0$. In the collisional limit, we distinguish between neutrinos scattering more rapidly than antineutrinos ($\Gamma_0>\overline{\Gamma}_0$) and the opposite ($\overline{\Gamma}_0>\Gamma_0$). The numerical values of $\omega_0$, $\Gamma_0$, and $\overline{\Gamma}_0$ chosen for different examples will be provided directly in Figs.~\ref{fig1}--\ref{fig4}, where we show the numerical results.

We always solve numerically the dispersion relation in Eq.~\eqref{eq:general_dispersion_relation}, starting with the $k=0$ solutions for $\omega\gg \tomegaE, \Gamma_E$, that we discussed in Sec.~\ref{sec:superluminal} (gapped modes) and Sec.~\ref{sec:gapless} (gapless ones). For the gapped modes, this approximation is always good, since by definition $\mathrm{Re}(\omega)\sim \mu \epsilon \gg \tomegaE, \Gamma_E$. For the gapless ones, this approximation is suitable only in the near-symmetric case ($\epsilon \ll 1$), where the eigenfrequencies are enhanced by $\epsilon^{-1}$. For the asymmetric case, we do not show the gapless solutions because
they will be seen to be phenomenologically less motivated.

From a numerical perspective, beginning with the $k=0$ solution, we can find those at larger $k$ through continuity. Specifically, we use a numerical root finding algorithm that uses the solution at a given $k$ as a starting point for finding the solution at the next $k$ value. In this process, we encounter the general difficulty that at large $k$, the modes closely approach the light cone, where the integrals become highly singular. To avoid this problem, when the values of $|\mathrm{Re}(\omega)|$ and $|k|$ become too close, we use the approximate form of the dispersion relation in Eqs.~\eqref{eq:dispersion_logarithmic_finite_k} and~\eqref{eq:dispersion_logarithmic_finite_k-B}. Solutions from the two regimes match very well, since these equations are excellent approximations near the light cone.

\subsection{Near-symmetric case}

\begin{figure}[ht]
\includegraphics[width=\textwidth]{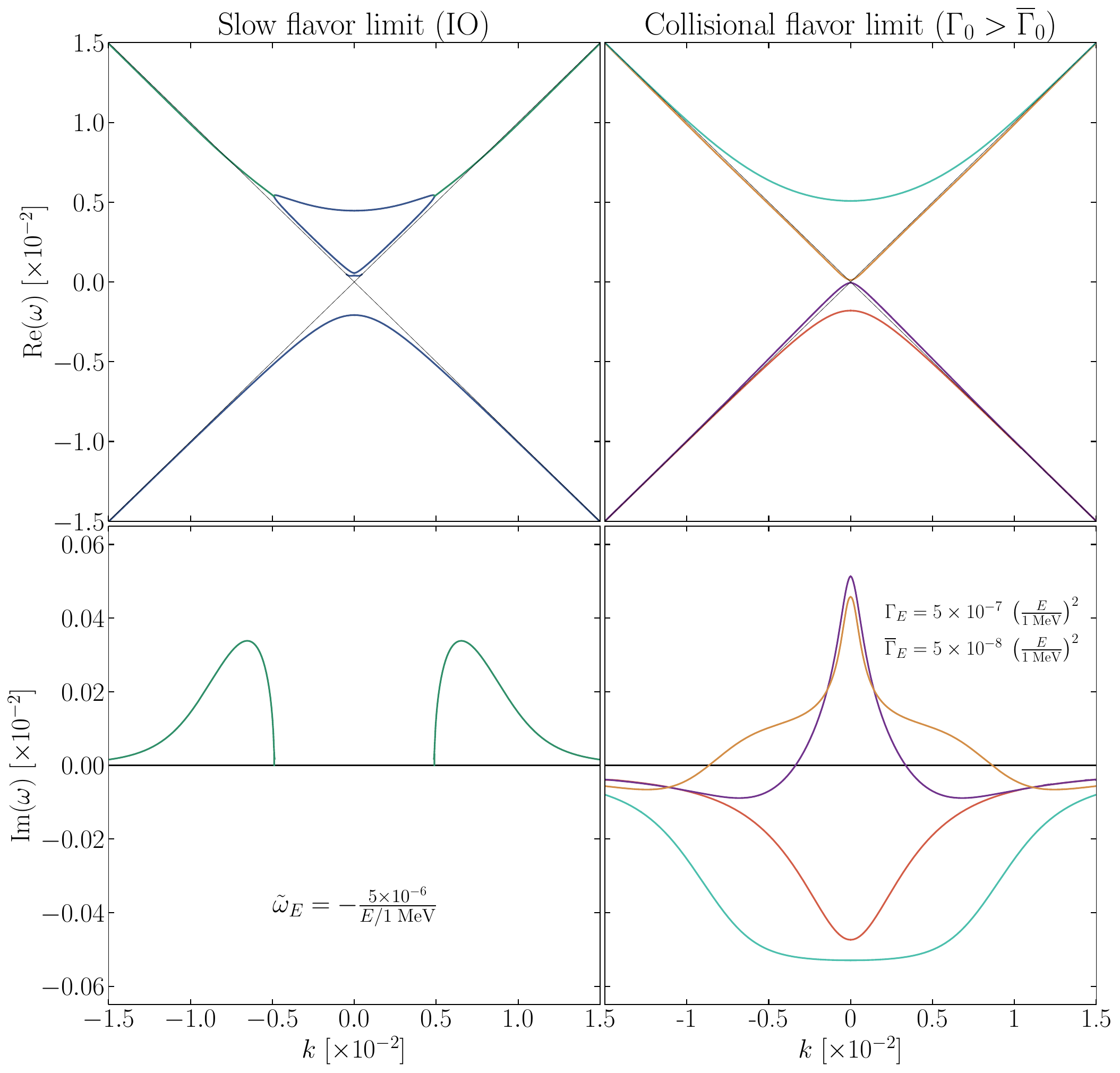}
\caption{Collective modes in the near-symmetric case
($\epsilon\ll1$), showing the real (top panels) and imaginary (bottom panels) part of the eigenfrequency as a function of wavenumber. Left panels: slow limit for IO, blue real-valued modes, green unstable ones. Right panels: collisional limit for $\Gamma_E>\overline{\Gamma}_E$, colors to differentiate among branches.
}\label{fig1}
\vskip12pt
\end{figure}

Beginning with the near-symmetric case ($\epsilon\ll1$), Fig.~\ref{fig1} shows the real and imaginary parts of the eigenfrequency $\omega$ as a function of $k$. We consider only IO for the slow limit, and $\Gamma_E>\overline{\Gamma}_E$ for the collisional one. We ignore the branch of subluminal gapped modes, which are Landau-damped already in the fast regime since there is no angular crossing.

\begin{figure}
\includegraphics[width=\textwidth]{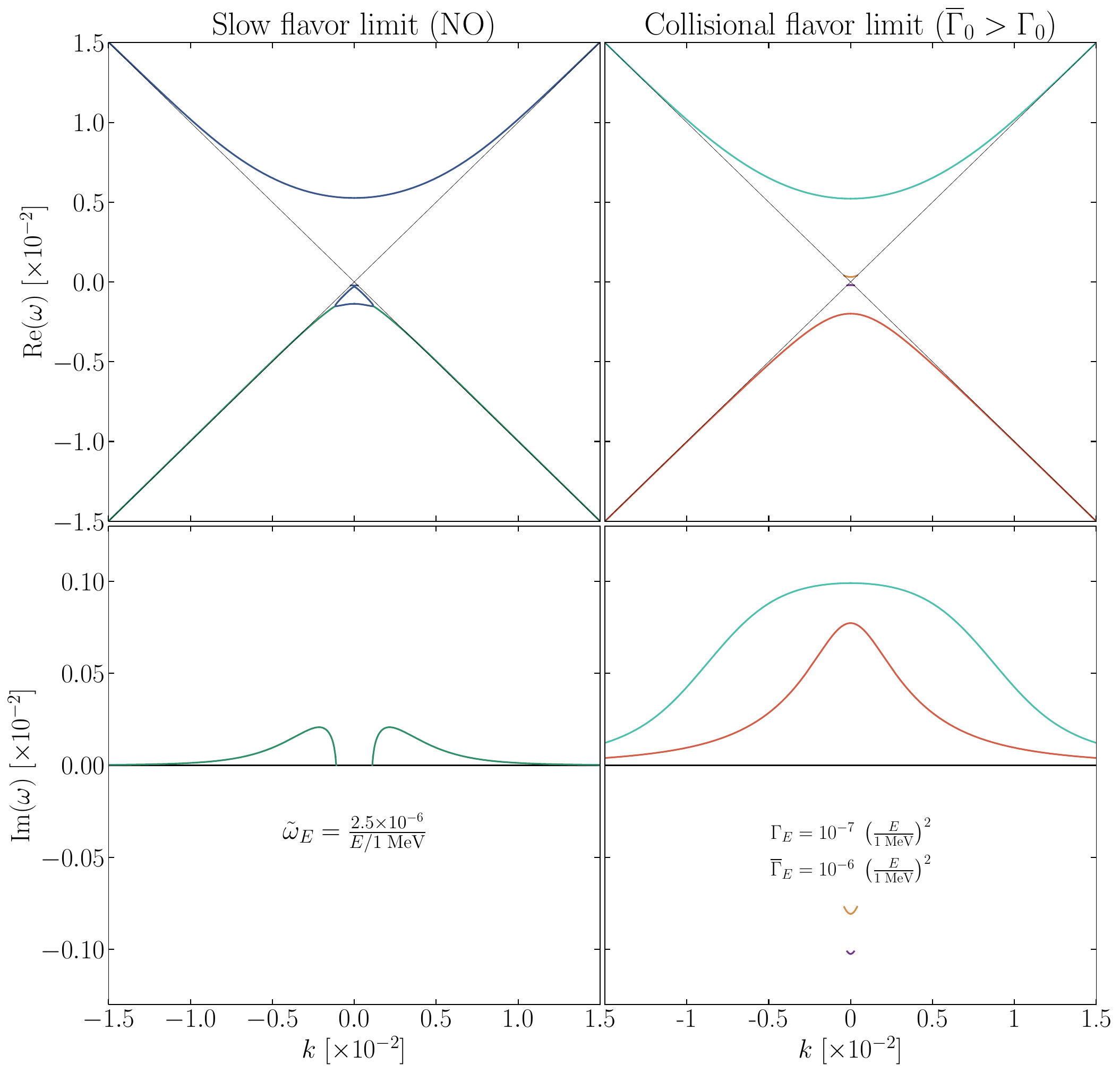}
\caption{Same as Fig.~\ref{fig1}, switching to NO in the slow case and to $\Gamma_E<\overline{\Gamma}_E$ in the collisional case.
}\label{fig2}
\end{figure}

For the slow limit, the instability pattern is very similar to the monochromatic case discussed in Ref.~\cite{Fiorillo:2024pns}. Typically $k\sim 10^{-2}$, in agreement with $\mu=1$, since here we used $\epsilon\sim 0.01$, the relative difference between overall neutrino and antineutrino densities. At $k=0$, the distinction between gapped and gapless modes is evident: all branches are real (blue), and two of them have $\mathrm{Re}(\omega)\sim \mu \epsilon$ while the other two have $\mathrm{Re}(\omega)\sim \omega_0/\epsilon$. The gapless modes are here also stable because, as discussed in Sec.~\ref{sec:gapless}, because $\mathrm{Re}(\omega)$ is much larger than a typical $\tomegaE$ in the bulk of the distribution. There are only negligibly few particles that satisfy the wave-particle resonance condition $\omega=\pm \tomegaE$ at $k=0$. The general approximations found in the previous sections for $\omega\gg \tomegaE$ are therefore very good for both gapped and gapless modes. 

For larger $k$, the lower gapped branch always remains real valued. One of the gapless branches exists only in the vicinity of $k=0$, and disappears when it touches the light cone.

In contrast, the other gapless and upper gapped branch merge to form an unstable near-luminal branch (green), one for positive and one for negative $k$. As anticipated, the order of magnitude of the growth rate is $\mathrm{Im}(\omega)\sim \omega_0/\epsilon$. While Fig.~\ref{fig1} suggests these unstable branches to extend to $k\to\pm\infty$, we have seen in Sec.~\ref{sec:near_luminal} that this happens only for monochromatic distributions, whereas here, the solution disappears at some large $k$. However, since $\epsilon\ll 1$, the monochromatic approximation holds up to very large $k$ because $\mathrm{Im}(\omega)\gg \tomegaE$, while for $\epsilon\sim 1$, the disappearance is abrupt and visible. We also note that these unstable modes, being superluminal, also have complex conjugates which are damped and not shown here.

Notice an apparent contradiction with our earlier theoretical argument in Sec.~\ref{sec:near_luminal}, where we concluded that in the slow limit, truly real-valued modes do not exist, since formally all waves always resonate with {\em some\/} neutrinos of particularly small energies and concomitant large $\tomegaE$. Here, however, the blue branches of Figs.~\ref{fig1} and \ref{fig2} appear to be real-valued. This paradox is only an optical illusion caused by very different scales. The blue branches do have invisibly small imaginary parts, which become sizable only close to the light cone, and are otherwise negligible compared with the much larger $\mathrm{Im}(\omega)\sim \tomegaE/\epsilon\sim 10^{-4}$ that are visible in Fig.~\ref{fig1} and~\ref{fig2}.

For the collisional limit, as expected, at $k=0$ we have two damped gapped modes and two unstable gapless ones. Once more, the approximate eigenvalues for $\omega\gg \Gamma_E$ hold very well. For larger $k$, all of these branches asymptote to the light cone, and at very large $k$, the gapless and gapped modes become degenerate (the purple band merges with the red one, while the orange band merges with the teal one). Overall, these results confirm that the collisional instabilities are concentrated around $k=0$ and are gapless, although the instability survives even at $k\sim \mu \epsilon$, where it becomes near-luminal.

In Fig.~\ref{fig2} we reverse cases: NO
for the slow limit and $\Gamma_E<\overline{\Gamma}_E$ for the collisional one, allowing us to test the earlier predictions about the nature of the eigenmodes. As expected,
in the slow limit, the change of mass ordering switches the unstable modes from the upper to the lower light cone, a behavior previously found for the monochromatic case~\cite{Fiorillo:2024pns}, and extended here to non-monochromatic distributions. However, we had to choose a smaller value for $|\omega_0|$, since for $\omega_0=10^{-5}$ the two unstable branches would merge to an unstable one already at $k=0$. This behavior is explained by the ``gap''  (the eigenfrequency at $k=0$) of the lower branch in the fast limit being significantly smaller than for the upper branch. Therefore a lower vacuum frequency is sufficient to significantly alter it.

In the collisional limit, the gapped and gapless branches switch their unstable and damped nature, as expected. The gapped branches are now unstable, while the gapless ones become damped, although they only exist close to $k=0$, and disappear when they touch the light cone. There is a peculiar difference compared to Fig.~\ref{fig1}, where the growing gapless modes passed through the light cone and survived up to large $k$. The difference is induced by causality: the dispersion relation is completely analytical for $\mathrm{Im}(\omega)>0$, whereas for $\mathrm{Im}(\omega)<0$, as mentioned several times, is defined as an analytical continuation. Hence, when the integrand functions $G(E,v)$ and $\oG(E,v)$ pass through a discontinuity at $v=\pm 1$, the dispersion relation becomes non-analytical. This explains why modes with $\mathrm{Im}(\omega)<0$ disappear when touching the light cone, whereas modes with $\mathrm{Im}(\omega)>0$ can pass freely through it. This effect was already identified in the context of fast instabilities in our previous works on fast and slow instabilities~\cite{Fiorillo:2024bzm, Fiorillo:2024uki, Fiorillo:2024pns}. The growth rate of the unstable gapped modes is $\mathrm{Im}(\omega)\sim \Gamma_0/\epsilon$, again in agreement with our qualitative expectations.

\subsection{Asymmetric case}

\begin{figure}
\includegraphics[width=\textwidth]{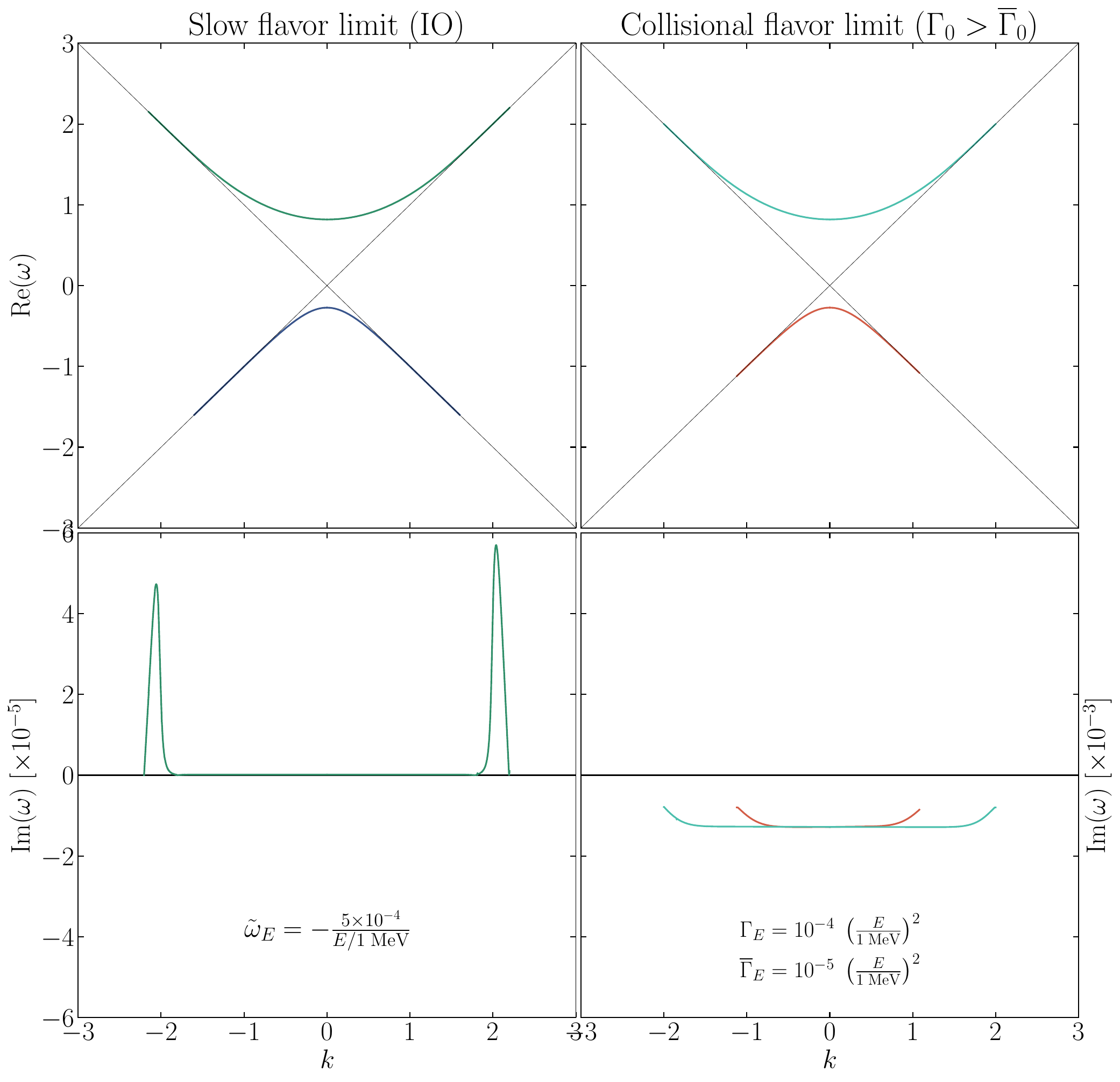}
\caption{Same as Fig.~\ref{fig1}, now for the asymmetric case ($\epsilon\sim1$).}\label{fig3}
\end{figure}

Turning now to the asymmetric case ($\epsilon\sim1$), Fig.~\ref{fig3} is analogous to Fig.~\ref{fig1} in that it shows collective modes for IO and the collisional ones for $\Gamma_E>\overline{\Gamma}_E$. Once more, we avoid showing the fast Landau-damped modes and now also avoid the gapless modes. We immediately observe that the typical wavenumber scale is $k\sim1$, since $\mu=1$ and $\epsilon\sim 1$. The general properties are similar to the near-symmetric case of Fig.~\ref{fig1}, yet present differences that are instructive to notice.

In the slow limit, the lower branch remains stable, while the upper one exhibits the usual double-hump structure. This mode remains unstable all the way to $k=0$, as expected from our arguments in Sec.~\ref{sec:near_luminal}, although the growth rate decreases as $k\to 0$ since the fraction of neutrinos that fulfills the resonance condition $\omega-vk\pm\tomegaE=0$ becomes smaller. We notice also that in this case the lower branch, shown in blue, actually must exhibit Landau damping. This is indeed true, but to track the precise value of the damping rate requires significant numerical precision in the energy integration. Since these modes are anyway Landau-damped, we do not show their damping rate. The most interesting feature, different from the near-symmetric case and from the monochromatic case studied in Ref.~\cite{Fiorillo:2024pns}, is that the unstable branches terminate abruptly at a small $k$. We had anticipated this behavior, as in a non-monochromatic setup there cannot be unstable solutions for $k\to \infty$.

Notice also that the two unstable branches are slightly asymmetrical, a behavior caused by the small anisotropy of our distribution. For $v=1$, the local value $\epsilon_v$ is slightly smaller than for $v=-1$, due to the excess of antineutrinos moving forward. Therefore, the right hump, corresponding to flavomons moving with a phase velocity $v_{\rm ph}=1$, is sensitive to a smaller value of $\epsilon_v$ and therefore has a slightly larger growth rate than the right hump.

Slow gapless modes are Landau-damped, as discussed in Sec.~\ref{sec:gapless}. They are difficult to obtain numerically as there is no good starting point for finding the solution, and furthermore, since they are Landau-damped, the integral must include the contribution from the deformed contour as $\mathrm{Im}(\omega)<0$. Since we know these solutions to be damped, we do not search for them numerically; see also our comments in Sec.~\ref{sec:discussion} on the physical relevance of gapless modes. As a sanity check, we have still tested that for $k=0$, where the integrals are particularly simple, the gapless solutions are indeed damped.

\begin{figure}
\includegraphics[width=\textwidth]{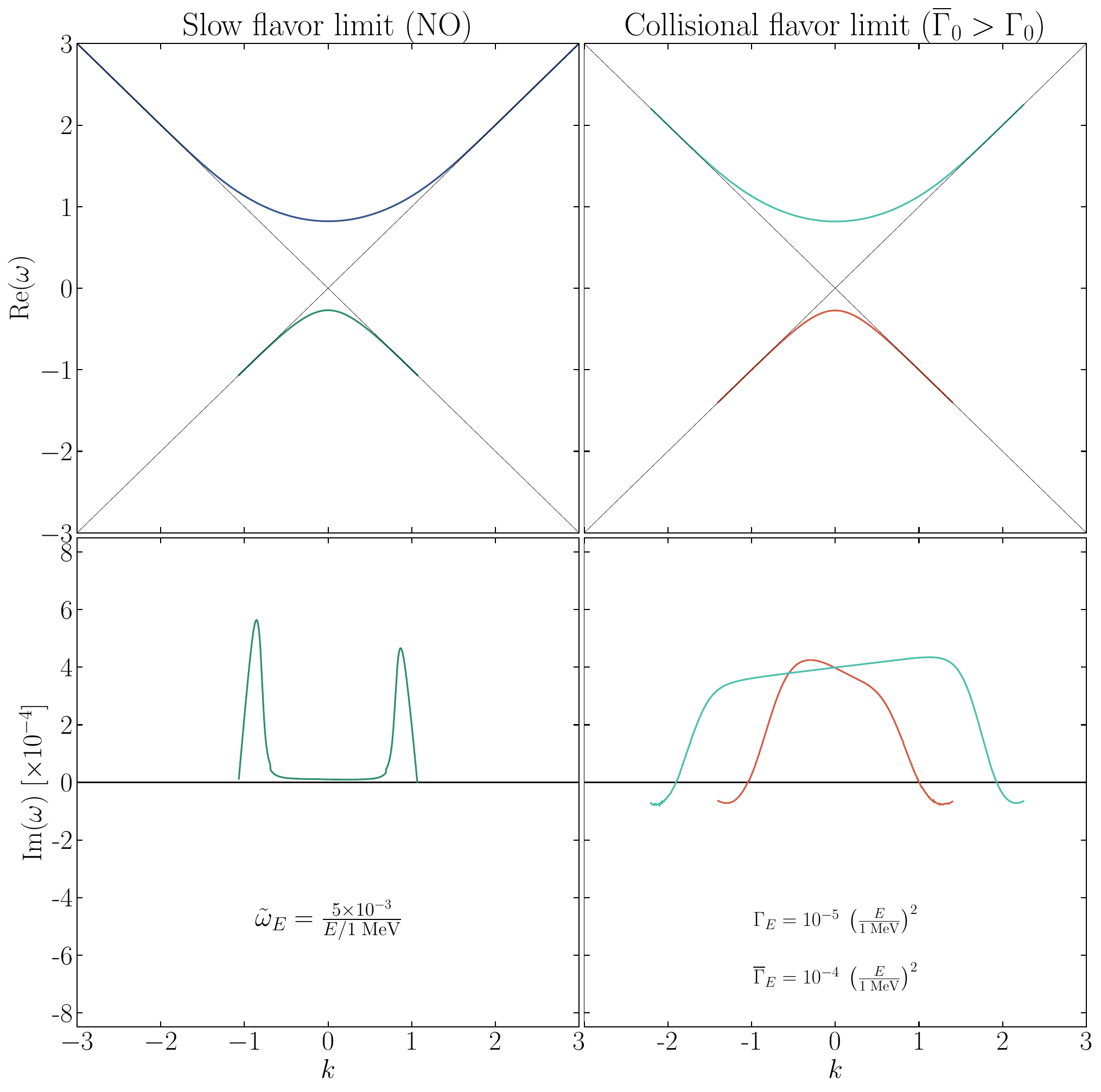}
\caption{
Same as Fig.~\ref{fig2}, now for the asymmetric case ($\epsilon\sim1$), and the same as Fig.~\ref{fig3} after switching to NO and $\Gamma_E<\overline\Gamma_E$.
}\label{fig4}
\end{figure}

In the collisional limit, the gapped modes start damped at $k=0$, as expected for this regime, and then approach the light cone at larger $k$ while their damping rate decreases. When these modes touch the light cone, they disappear. The gapless collisional modes might be unstable; however, as discussed in Sec.~\ref{sec:gapless}, for sufficiently large $\epsilon$, they might disappear altogether. This is indeed the case for our example distribution; we have explicitly sought a solution around $k=0$, without finding one. To verify that such solution does not exist, we have explicitly tested an identical energy and angular distribution with a smaller $\epsilon$, by taking $\overline{\mathcal{N}}\simeq \mathcal{N}$, for which the two gapless unstable solutions exist and can be found as described in Sec.~\ref{sec:gapless}. We have gradually reduced $\overline{\mathcal{N}}$ until it reaches the value of the asymmetric distribution. We find, as expected, that both gapless solutions disappear before this happens, confirming that there are no unstable gapless modes.

Finally, Fig.~\ref{fig4} switches to NO and $\Gamma_E<\overline\Gamma_E$ and therefore is the asymmetric ($\epsilon\sim 1$) version of Fig.~\ref{fig2}, which pertains to $\epsilon\ll 1$. The results confirm what we have already understood: in the slow limit, the unstable modes have moved to the lower, rather than upper, light cone. In this case, it is the left hump that is slightly larger, since it now corresponds to flavomons moving forward with phase velocity $v_{\rm ph}=+1$. In the collisional limit, the gapped modes are unstable
rather than stable. Regarding the gapless modes, which are not shown here, in the slow limit they are Landau-damped as discussed earlier. In the collisional limit, the gapless modes might be unstable; however, as discussed in Sec.~\ref{sec:gapless}, for sufficiently large $\epsilon$, they might disappear altogether. We have followed the same procedure as for the case of Fig.~\ref{fig3} to verify that for our example distribution, there are indeed no unstable gapless modes.

\subsection{Disappearance of collisionally unstable gapless modes}

\begin{figure}[b]
    \includegraphics[width=\textwidth]{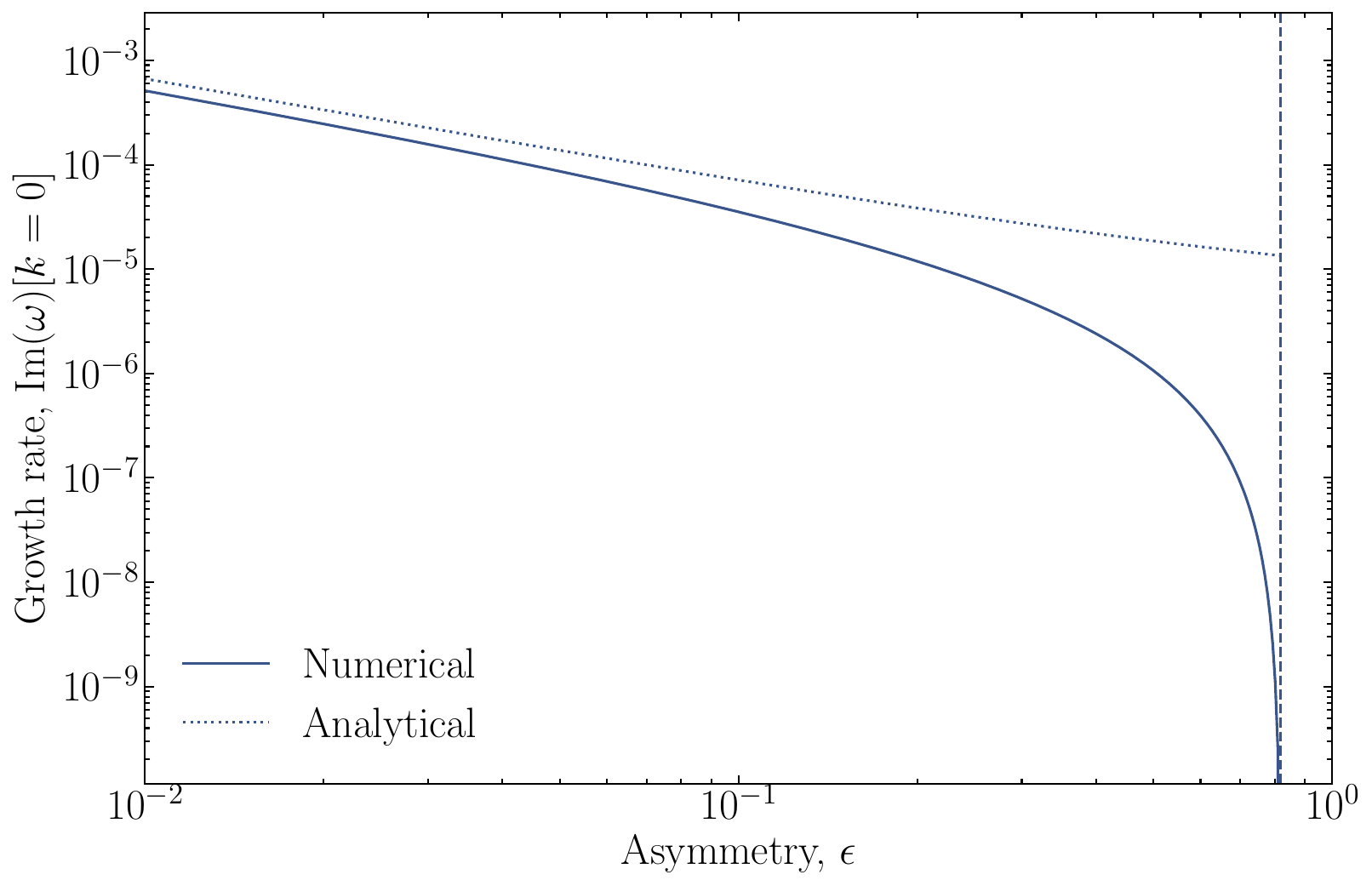}
    \caption{Growth rate of collisionally unstable gapless modes as a function of the asymmetry $\epsilon$ of the distribution. Numerical solution (solid) compared with theoretical prediction (dotted) of Eq.~\eqref{eq:gapless_near_symmetric}, and expected position of disappearance of the gapless mode (dashed) from Eq.~\eqref{eq:condition_disappearance_collisional}.}\label{fig5}
\end{figure}

We have examined two cases, a near-symmetric one ($\epsilon\ll 1$) for which gapless modes can be unstable, and an asymmetric one ($\epsilon\sim1$) for which such modes have disappeared altogether. It is instructive to explicitly show this transition, and in particular its agreement with our analytical prediction. The simplest scenario is a purely isotropic case ($\epsilon_v=0$ in Eq.~\ref{eq:distributions}) with $T=\overline{T}=1$~MeV. We also focus on $k=0$, which is the mode from which the gapless unstable modes emerge as shown in Fig.~\ref{fig1}. For $k=0$ and isotropy, the dispersion relation factorizes in a monopole one ($I_0=1$) and a quadrupole one ($I_2=-1$). We solve the former numerically, while we vary the asymmetry by taking
\begin{equation}
    \mathcal{N}=\frac{1+\epsilon}{2}
    \quad\mathrm{and}\quad
    \overline{\mathcal{N}}=\frac{1-\epsilon}{2}.
\end{equation}
Finally, we take the collision rates $\Gamma_0=10^{-6}$ and $\overline{\Gamma}_0=10^{-7}$. Since $\Gamma_E>\overline{\Gamma}_E$ and $\epsilon>0$, this is a regime where collisionally unstable gapless modes should show up.

Based on our arguments in Sec.~\ref{sec:gapless}, for $\epsilon\ll1$, the growth rate is approximately given by Eq.~\eqref{eq:gapless_near_symmetric}. For our distributions, we find
\begin{equation}
    D_0=\epsilon
    \quad\mathrm{and}\quad
    \Sigma_0=12\left[\Gamma_0\frac{1+\epsilon}{2}+\overline{\Gamma}_0\frac{1-\epsilon}{2}\right],
\end{equation}
where the factor $12$ comes from averaging the $E^2$ dependence of the collision rate over the energy distributions. In addition, from Eq.~\eqref{eq:condition_disappearance_collisional}, we expect the collisionally unstable gapless mode to disappear at $\mathcal{N}/\overline{\mathcal{N}}=\Gamma_0/\overline{\Gamma}_0=10$, that is, at $\epsilon=9/11$.

Figure~\ref{fig5} shows the resulting growth rate, together with our theoretical expectation. At small $\epsilon$, these match extremely well; the small offset is caused by our approximation in Eq.~\eqref{eq:gapless_dispersion_relation} of neglecting 1 on the right-hand side, which otherwise would produce an even more accurate expression. As $\epsilon$ grows, the numerical growth rate diverges downward from the linear approximate one, until it drops to 0 exactly at $\epsilon=9/11$, the value predicted by our exact criterion in Eq.~\eqref{eq:condition_disappearance_collisional}. This behavior confirms our general predictions and understanding of the collisionally unstable gapless modes.

\subsection{Narrow slow instabilities}\label{sec:narrow_slow_numerical}

As we have discussed in Sec.~\ref{sec:near_luminal}, if the energy distribution is not particularly symmetric among neutrinos and antineutrinos, or more generally, if the energy-integrated lepton number difference along a certain direction is not particularly small, the growth rate of slow instabilities is not enhanced. In these cases, on the contrary, it can be smaller than, or comparable with, the typical vacuum frequencies. As one such example, for generality we consider an instability caused by a crossing of $\nu_e$ and $\nu_x$ at a finite energy instead of the crossing between $\nu_e$ and $\overline{\nu}_e$, which can be interpreted as a spectral crossing at zero energy. Thus, we choose the $\nu_e$ and $\nu_x$ energy and angular distribution in the form
\begin{equation}\label{eq:slow_distribution}
    f_{e}=\frac{1}{2}\mathcal{N}_{e}\left(\frac{E}{T_{e}}\right)^2 \frac{e^{-E/T_{e}}}{2T_{e}}
    \quad\textrm{and}\quad
    f_\mu=\frac{1}{2}\mathcal{N}_{\mu}\left(\frac{E}{T_{\mu}}\right)^2 \frac{e^{-E/T_{\mu}}}{2T_{\mu}}.
\end{equation}
The numerical values are chosen as $\mathcal{N}_{e}=0.98$, $\mathcal{N}_{\mu}=0.02$ (normalized to $\mathcal{N}_{e}+\mathcal{N}_{\mu}=1$ so that $\mu=1$), $T_{e}=1\, \mathrm{MeV}$, and $T_{\mu}=3\,\mathrm{MeV}$. As usual, the evolution of flavor conversions is ruled by $G(v,E)=f_e(v,E)-f_\mu(v,E)$. The larger temperature of the $\nu_x$ lets them dominate at large energies, even though they are largely subdominant at energies of a few MeV.  For this case, we choose $\omega_0=10^{-2}$, which is substantially larger than the typical values we had to assume for the previous cases, for which the neutrino-antineutrino asymmetry was always chosen quite small and therefore a very small value of $\omega_0$ was needed to remain in the regime $\tomegaE\ll \mu \epsilon^2$. With this energy distribution, instead, $\epsilon$, interpreted as the overall percentage of DLN, is of order 1, and therefore $\tomegaE\sim 10^{-2}$ is already in the correct qualitative regime of instability.

\begin{figure}[ht]
    \includegraphics[width=\textwidth]{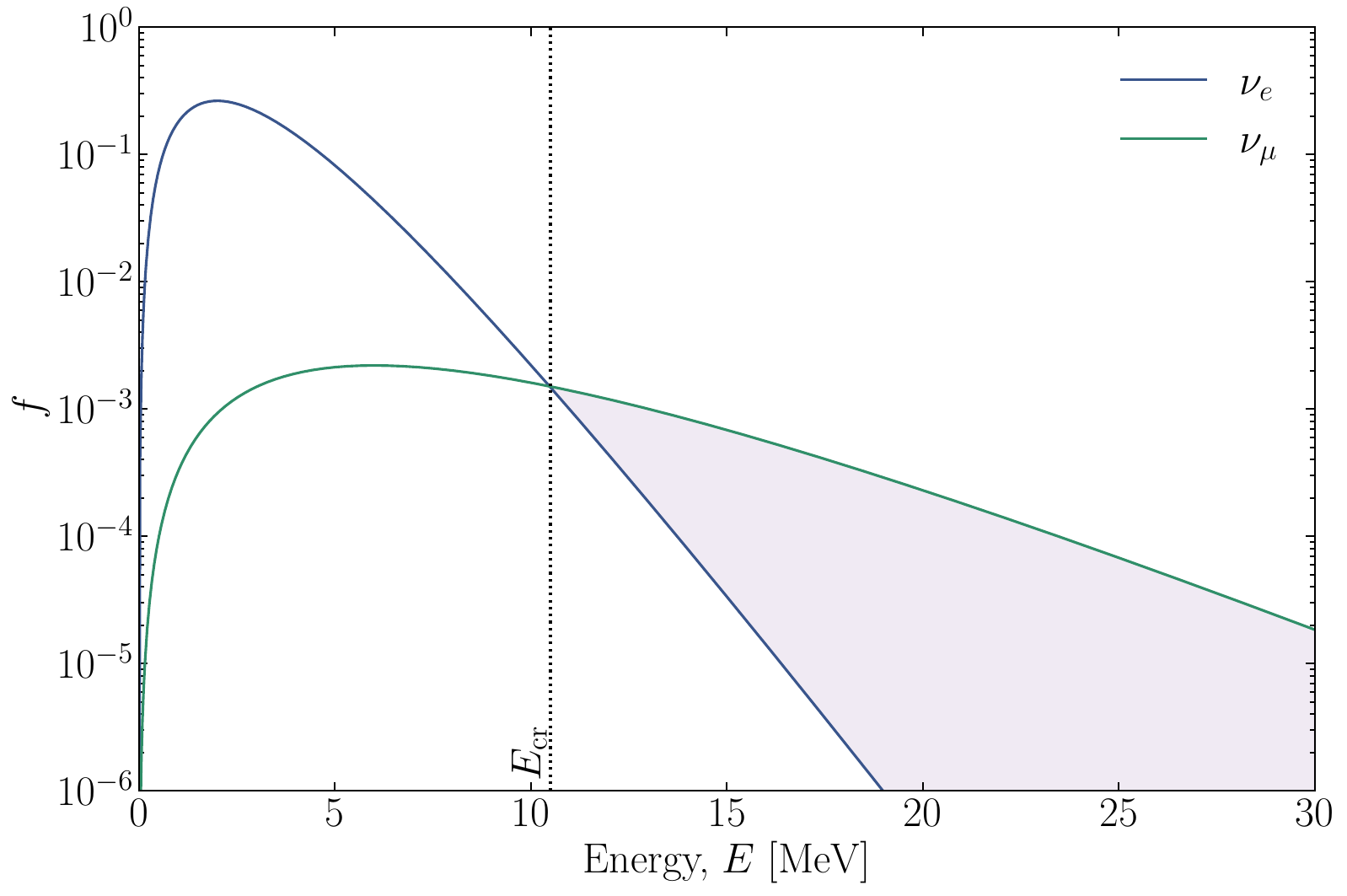}
    \caption{Energy distribution of $\nu_e$ and $\nu_\mu$ for our benchmark case of narrow slow instability defined by Eq.~\eqref{eq:slow_distribution}. The angular distribution is isotropic for both species.}\label{fig:slow_distribution}
\end{figure}

The energy distributions thus defined are shown in Fig.~\ref{fig:slow_distribution}, where our logarithmic axis enhances visibility of the crossing region. There is a clearly dominant $\eln$ region at low energies caused by $\nu_e$, whereas above $E_{\rm cr}$, highlighted by a black dotted line, $\nu_\mu$ dominate. We call them the flipped neutrinos, a nomenclature introduced in Refs.~\cite{Fiorillo:2024bzm, Fiorillo:2024uki}, which means that these neutrinos have flipped $\eln$ relative to the dominant population and are responsible for the instability. The flipped $\eln$ region is highlighted by purple shading, so the total number of neutrinos integrated in that area is $n_{\nu, \rm fl}$.

\begin{figure}
    \includegraphics[width=\textwidth]{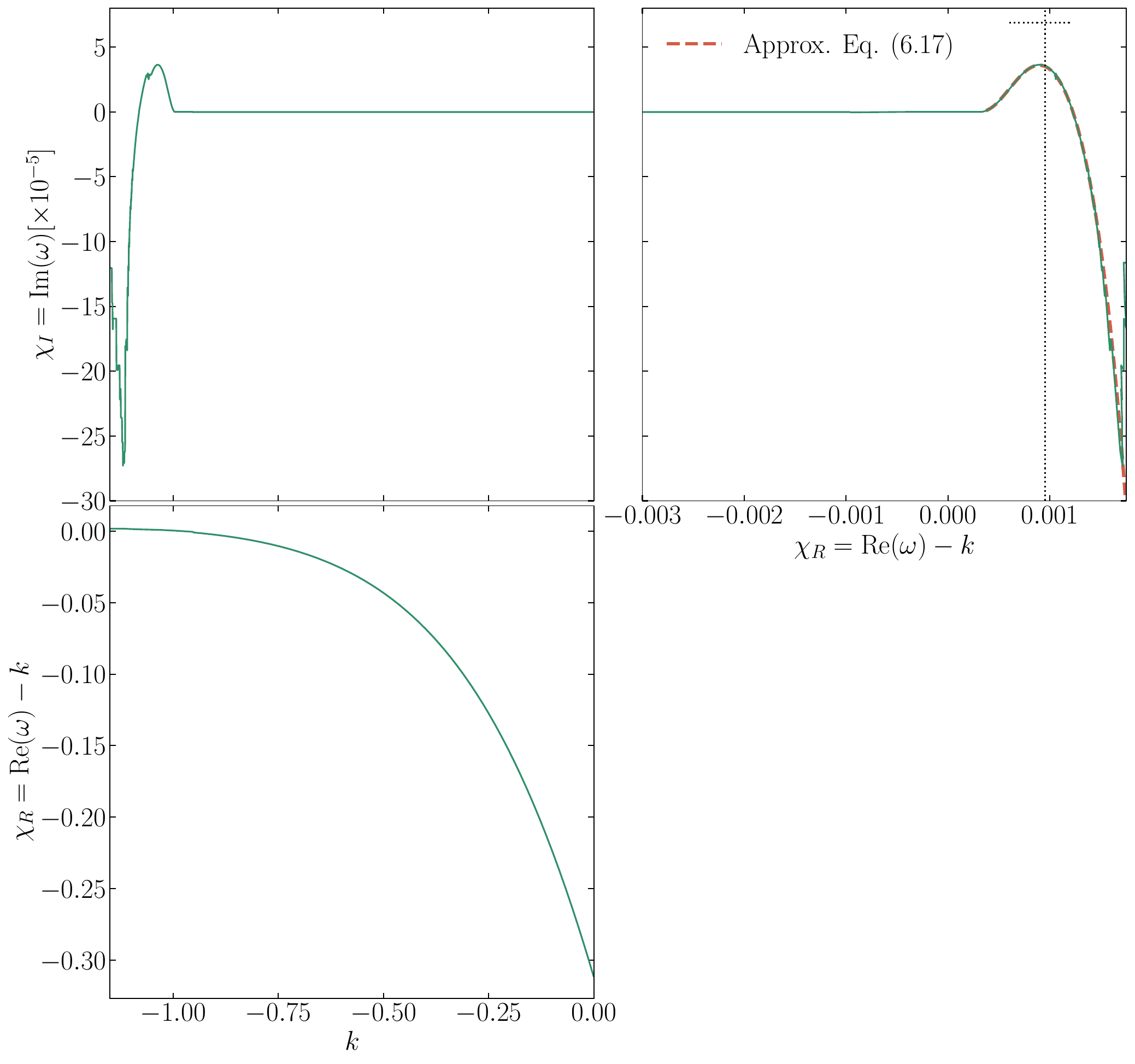}
    \caption{Collective slow-unstable mode for a narrow slow instability. We show the imaginary (top) and real (bottom) part of the eigenfrequency, shifted by $k$, as a function of wavenumber $k$ (left) and of the shifted real part of the eigenfrequency (right). We also show the approximate growth rate obtained using Eq.~\eqref{eq:approximate_growth_rate_slow_resonant} to compare it with the numerical result.}\label{fig:narrow_slow}
\end{figure}

We focus on NO, neglect collisions, and consider the lower branch with $\mathrm{Re}(\omega)<0$; as we have seen in Figs.~\ref{fig2} and~\ref{fig4}, this is the branch that becomes unstable in NO. The corresponding imaginary part of the eigenfrequency is shown in Fig.~\ref{fig:narrow_slow}. At $k\to 0$ the mode is superluminal and therefore cannot interact with any neutrino, so it has no growth rate. When $\chi_R=\mathrm{Re}(\omega)-k$ becomes sufficiently small, the flavomons are kinematically allowed to interact with some of the neutrinos, satisfying the resonance condition $\mathrm{Re}(\omega)-k-\tomegaE=0$ for some energy. The growth rate becomes positive, producing a small bump, because at relatively small and positive values of $\chi_R$, the contribution comes only from neutrinos with very large energies, exceeding $\omega_0/\chi_R$, which are of $\nu_\mu$ type and therefore tend to produce the instability. (At negative values of $\chi_R$, due to the absence of antineutrinos, the imaginary part of the dispersion relation must vanish, see Eq.~\ref{eq:approximate_growth_rate_slow_resonant}.) 

As $\chi_R$ becomes larger, the numerator in Eq.~\eqref{eq:approximate_growth_rate_slow_resonant} acquires a contribution also from the $\nu_e$ dominating at low energies, which flip the sign of $\mathrm{Im}(\omega)$ and cause the growth rate to turn negative, which happens when
\begin{equation}
    \int_{\omega_0/\chi_R}^{+\infty}G(1,E)\,dE=0.
\end{equation}
We also report in Fig.~\ref{fig:narrow_slow} the analytical prediction of Eq.~\eqref{eq:approximate_growth_rate_slow_resonant}, which matches perfectly with the numerical result in the limit of small growth rate. Notice that at larger values of $\chi_R$, the growth rate turns negative and begins to show a scattered behavior; this is because at large~$\chi_R$, the neutrino-flavomon interactions sample neutrinos at lower energies $E\sim \omega_0/\chi_R$, where the discrete energy grid strongly shows up. Luckily, these modes are generally uninteresting because they are Landau-damped and certainly not unstable. Overall, we find extremely good agreement between the numerical results and our analytical formula for the growth rate in the narrow-slow-instability regime.

As a simpler and more intuitive guide to the properties of the instability, we consider the order of magnitude of the maximum growth rate according to Eq.~\eqref{eq:approximate_growth_rate_slow_resonant}, which is roughly reached when $\chi_R\simeq \tilde{\omega}_{E_{\rm cr}}$. Increasing $\chi_R$, the integral in the numerator receives contributions from the non-flipped neutrinos, which have the opposite $\eln$ and reduce the growth rate. Therefore, the value of $\chi_R=\omega-k$ at which the growth rate is maximum is $\chi_R\simeq \tilde{\omega}_{E_{\rm cr}}$, highlighted in Fig.~\ref{fig:narrow_slow} by a vertical dotted line. The agreement is very good. The typical growth rate at this position can be estimated in order of magnitude by approximating further Eq.~\eqref{eq:approximate_growth_rate_slow_resonant}; the numerator simply equals $n_{\nu, \rm fl}$, the number density of flipped neutrinos along the chosen direction. In the denominator, if the crossing is at very large energies, as expected for these weak-instability cases, we can assume $\chi_R\ll \tomegaE$ for the bulk of the distribution and neglect it altogether. Therefore, we find, as an order-of-magnitude estimate,
\begin{equation}
    \chi_I\sim \frac{\pi \omega_0 n_{\nu, \rm fl}}{\mathcal{E}_\nu}
  \quad\textrm{with}\quad
    \mathcal{E}_\nu= \int dE D(1,E)\,E,
\end{equation}
the latter being the $\eln$-energy density along the chosen direction. This estimate is reported as a horizontal dotted line in Fig.~\ref{fig:narrow_slow}, showing that indeed it correctly captures the order of magnitude of the maximum growth rate.

\section{Thermodynamics of collisional instabilities}\label{sec:thermodynamics}

Collisional effects, in our treatment, are limited to beta processes involving $\nu_e$ and $\overline\nu_e$, which are driven to equilibrium with the medium. The other flavors are assumed to interact only through neutrino-neutrino refraction and, therefore, typically do not reach thermal and chemical equilibrium. Consequently, the original proof of the H-Theorem in the presence of neutrino-neutrino refraction \cite{Sigl:1993ctk} no longer holds, as it assumed a final neutrino equilibrium state with a common chemical potential for all flavors. Here, we extend the proof to accommodate our more general case of several chemical potentials.

\subsection{Irreversibility from coarse graining vs.\ collisions}

Slow and fast instabilities share a common feature: they are both irreversible processes arising within an entropy-conserving system. In systems with only a few discrete degrees of freedom, the evolution is actually reversible. However, once a continuous distribution of energy and angle is considered, the instability becomes effectively irreversible, even though the neutrino entropy remains unchanged. This appearance of irreversibility in an apparently reversible system is well-known from plasma physics, where it signals the presence of fluctuations, usually rapidly varying in time and space, which are ``excluded'' from the averaged distribution. The average is performed here over scales comparable with the wavelength and period of the unstable modes, which are much shorter than the scales over which the original distribution varies. This coarse-graining procedure leads to a loss of information and produces irreversibility. In the context of collective flavor conversions, this argument was advanced by Johns \cite{Johns:2023jjt}, who was first to introduce the entropy for the space-time averaged neutrino distribution. He hypothesized that such entropy could only grow and suggested that a BBGKY hierarchy might in future be used for a proof.

We now turn to actually proving Johns' conjecture, but relying on a different argument. With the neutrino-flavomon formalism introduced in Ref.~\cite{Fiorillo:2025npi}, we can look at irreversibility from a different perspective. In this picture, the instability is caused by the decay $\nu_\mu\to\nu_e+\psi$ (or $\overline{\nu}_e\to \overline{\nu}_\mu+\psi$ if there are antineutrinos), and therefore one can immediately introduce a coarse-grained entropy satisfying the H-theorem
\begin{equation}
    S=S_\nu+S_\psi=S_{\nu_e}+S_{\nu_\mu}+S_\psi,
\end{equation}
with
\begin{equation}
    S_{\nu_e}=-\int \frac{d^3\bp}{(2\pi)^3}\Bigl[n_{\nu_e,\bp}\log n_{\nu_e,\bp}+(1-n_{\nu_e,\bp})\log(1-n_{\nu_e,\bp})\Bigr],
\end{equation}
where $n_{\nu_e,\bp}$ is the space-time averaged phase-space distribution (occupation number) for the electron neutrinos, and an identical expression holds for the muon neutrinos.
For the flavomons, in terms of their own average phase-space distribution $N_\bK$, we have
\begin{equation}
    S_\psi=-\int \frac{d^3\bK}{(2\pi)^3}\Bigl[N_\bK\log N_\bK-(1+N_\bK)\log(1+N_\bK)\Bigr].
\end{equation}
Since this species obeys a decay-like equation
\begin{eqnarray}
    \frac{\partial N_\bK}{\partial t}&=&\int \frac{d^3\bp}{(2\pi)^3}\int\frac{d^3\bp'}{(2\pi)^3}W(\bp\to\bp',\bK)
    \nonumber\\[1.5ex]
    &&{}\times
    \Bigl[n_{\nu_\mu,\bp}(1-n_{\nu_e,\bp'})(1+N_\bK)-n_{\nu_e,\bp'}N_\bK(1-n_{\nu_\mu,\bp})\Bigr],
\end{eqnarray}
and similarly for the neutrino species, one can prove explicitly that the total 
coarse-grained entropy $S_\nu+S_\psi$ must always increase. The concrete expressions for the decay probabilities $W(\bp\to\bp',\bK)$ are given in Ref.~\cite{Fiorillo:2025npi} in the approximation of weak instabilities.

Our coarse-grained neutrino entropy $S_\nu$ coincides with the one introduced earlier by Johns \cite{Johns:2023jjt}. At first sight, it appears that only the total neutrino plus flavomon entropy is actually bound to increase. However, intriguingly, in the quasi-classical regime $N_\bK\gg 1$, which is the only case of practical interest for flavor instabilities as detailed in Ref.~\cite{Fiorillo:2025npi}, the neutrino entropy alone must also increase. Indeed, by explicit computation we find
\begin{align}
\frac{dS_{\nu}}{dt} &=  \int \frac{d^3\bp}{(2\pi)^3} \frac{d^3\bp'}{(2\pi)^3} \frac{d^3\bK}{(2\pi)^3} \, W(\bp \to \bp', \bK) \\ \nonumber
&\quad \times \Bigl[ n_{\nu_\mu,\bp}(1 - n_{\nu_e,\bp'})(1 + N_\bK) - n_{\nu_e,\bp'} N_\bK (1 - n_{\nu_\mu,\bp}) \Bigr]  
\log \left( \frac{n_{\nu_\mu,\bp}(1 - n_{\nu_e,\bp'})}{(1 - n_{\nu_\mu,\bp}) n_{\nu_e,\bp'}} \right).
\end{align}
In the quasi-classical limit $N_\bK\gg 1$, we can now write
\begin{align}
    \frac{dS_\nu}{dt}&=\int \frac{d^3\bp}{(2\pi)^3} \frac{d^3\bp'}{(2\pi)^3} \frac{d^3\bK}{(2\pi)^3} \, W(\bp \to \bp', \bK) N_\bK \Bigl[n_{\nu_\mu,\bp}(1-n_{\nu_e,\bp'})-n_{\nu_e,\bp'}(1-n_{\nu_\mu,\bp})\Bigr]\nonumber\\[1ex]
    &\quad\times \log \left( \frac{n_{\nu_\mu,\bp}(1 - n_{\nu_e,\bp'})}{(1 - n_{\nu_\mu,\bp}) n_{\nu_e,\bp'}} \right),
\end{align}
an expression that is now positive-definite. Physically, a large occupation number of flavomons renders the flavor field classical. When spontaneous flavomon emission is neglected, the stimulated emission and absorption processes reduce to two effective conversions, $\nu_\mu \leftrightarrow \nu_e + \psi$, occurring at equal rates set by the flavomon background. This background acts as a classical bath and a sink of flavor, enabling efficient neutrino conversion $\nu_\mu \leftrightarrow \nu_e$, which in turn leads to the growth of neutrino entropy.

This calculation validates the conjecture of Ref.~\cite{Johns:2023jjt} and shows that in practice, the coarse-grained entropy of the neutrino distribution increases, due to the stimulated production of flavomons. On the other hand, this does not mean that the final state of flavor conversions must be one of complete thermal equilibrium, and in fact it generally is not; a well-known counter-example from plasma physics is the prototypical bump-on-tail instability, in which a bump of energetic electrons on top of a thermal electron distribution produces an instability with a growing number of plasmons \cite{schKT}. These plasmons lead to the disappearance of the bump, with the energy distribution flattening,
yet the final distribution is definitely not thermal. Thus, while the final state of kinetic instabilities has a higher entropy, both in the sense of $S_\nu$ and $S_\nu+S_\psi$, it is not generally the thermal state that maximizes these 
coarse-grained entropies.

Collisional instabilities, on the other hand, differ significantly in that neutrino entropy is not conserved even \textit{before} any form of coarse-graining. Therefore, while the picture of neutrinos decaying into flavomons formally still holds, it becomes intuitively less appealing, as neutrinos are no longer stable eigenstates. Fundamentally, collisional instabilities are not driven by an increase in neutrino-flavomon entropy, but rather by a decrease in the \textit{true} free energy of the system, without the need for any coarse-graining. Even a completely homogeneous system subject to a collisional instability will evolve into a state different from the initial one, as demonstrated by the slow-fast approximation scheme introduced in Ref.~\cite{Fiorillo:2023ajs}. To highlight this fundamental difference, we must explicitly determine the behavior of the free energy in the presence of collisions.

This was previously proven under the assumption that all flavors tend toward the same equilibrium distribution~\cite{Sigl:1993ctk}, as must occur when flavor is not conserved and sufficient time is available to reach \textit{true} equilibrium. However, over timescales short compared to those of flavor violation induced by mixing angles, and short compared to supernova diffusion timescales, neutrinos may still achieve an intermediate state of equilibrium in which flavor conversion is negligible and the chemical potentials of different flavors are not identical. We recall that collective flavor evolution alone does not cause a global violation of lepton number; such violation is only achieved through the mixing angles, which are neglected in the linear theory, although they are assumed to provide seeds for instabilities. In a supernova core, a large amount of electron lepton number is trapped, while muon and tau lepton numbers are absent. This situation can change only through true flavor conversion and transport to the surface. Therefore, we must generalize the original proof to the case where the chemical potentials of different flavors are distinct.

\subsection{Evolution of the free energy}

To avoid cluttering the calculations with unnecessary complications, we will revert here to a homogeneous and isotropic system, where the neutrino modes depend only on energy and time. The more general case is a straightforward extension. The energy-dependent density matrix $\rho(E)$ evolves through a kinetic equation schematically of the form
 \begin{equation}\label{eq:generic_eom}
 \frac{\partial \rho(E)}{\partial t}=-i\bigl[\sH(E),\rho(E)\bigr]+\frac{\bigl\{\tilde{\sA}_E,\rho^0(E)-\rho(E)\bigr\}}{2}.
 \end{equation}
Here $\sH(E)$ is the collisionless Hamiltonian of the neutrino dynamics, and $\tilde{\sA}_E=\sP_E+\sA_E$ is the enhanced absorption rate for neutrinos with energy $E$; these are all matrices in flavor space. Moreover, the equilibrium density matrix is
\begin{equation}
\rho^0(E)=\frac{1}{\exp\Bigl[\frac{E-\mu}{T}\Bigr]+1},
\end{equation}
where $\mu$ is a $3{\times}3$ matrix of chemical potentials. In the weak interaction basis, it would be $\mu={\rm diag}(\mu_{\nu_e}, \mu_{\nu_\mu}, \mu_{\nu_\tau})$, noting that these chemical potentials characterize the medium, not the true neutrino distribution, i.e., these are the chemical potentials the neutrinos would attain in beta equilibrium with the medium.

In our physical situation, allowing the medium to provide both energy and lepton number to the neutrino ensemble, the relevant thermodynamic potential is the Landau free energy or grand potential, 
\begin{equation}\label{eq:grandpotential}
\Phi=U-T S-\int dE\,\mathrm{Tr}\left[\mu \rho(E)\right],
\end{equation}
to be compared with Eq.~(3.11) of Ref.~\cite{Sigl:1993ctk}.
The total energy is
\begin{equation}
U=\int dE\, E \mathrm{Tr}\left[\rho(E)\right]
\end{equation}
and the total entropy is
\begin{equation}
S=-\int dE\,\mathrm{Tr}\left[\rho(E) \log\rho(E)+(1-\rho(E)) \log(1-\rho(E))\right].
\end{equation}

After differentiating all terms in Eq.~\eqref{eq:grandpotential}, the time derivative of the free energy is found to be a relatively simple expression
\begin{equation}
\frac{\partial \Phi}{\partial t}=T\int dE\, \mathrm{Tr}\left[\frac{\partial \rho(E)}{\partial t}\left(\log\frac{\rho(E)}{1-\rho(E)}-\log\frac{\rho^0(E)}{1-\rho^0(E)}\right)\right],
\end{equation}
where we can now replace the equations of motion Eq.~\eqref{eq:generic_eom}. The result is most easily interpreted by introducing the projector $P_i(\rho(E))$ over the $i$-th eigenstate of the matrix $\rho(E)$, with eigenvalues $\rho_i(E)$. In this way the time derivative is found to be
\begin{equation}
\frac{\partial \Phi}{\partial t}=\frac{T}{2}\int dE\sum_{i,j}\mathrm{Tr}\left[P_i(\rho) \tilde{\sA}_E P_j(\rho^0)+P_j(\rho^0)\tilde{\sA}_E
P_i(\rho)\right](\rho^0_j-\rho_i)\left[\log\frac{\rho_i}{1-\rho_i}-\log\frac{\rho^0_j}{1-\rho^0_j}\right]\!,
\end{equation}
where the dependence on $E$ of the $\rho$ matrices is understood. Moreover, the basis in which $\rho^0$ is diagonal, which defines the projector $P_j(\rho^0)$, must also diagonalize $\tilde{\sA}_E$, i.e., the weak interaction basis. Therefore, the expression in the first parenthesis becomes $2\tilde{\sA}_{E,j}\mathrm{Tr}\left[P_i(\rho) P_j(\rho^0)\right]$, which is positive-definite. In addition, the entire expression can be rewritten in the form
\begin{equation}
\frac{\partial \Phi}{\partial t}=\frac{T}{2}\int dE\sum_{i,j}\mathrm{Tr}\left[P_i(\rho) \tilde{\sA}_E P_j(\rho^0)+P_j(\rho^0) \tilde{\sA}_E P_i(\rho)\right]\rho^0_j(1-\rho_i)(1-x)\log x,
\end{equation}
with $x=\rho_i (1-\rho^0_j)/\rho^0_j(1-\rho_i)$. Since $(1-x)\log x<0$ for any $x$, the derivative of the free energy has been proven to be negative. 

This completes the proof of Ref.~\cite{Sigl:1993ctk} under our more general assumption of different chemical potentials for different flavors. It shows that collisions reduce the free energy, and that collisional instabilities are therefore intrinsically irreversible, even without coarse-graining over small scales. The ability of collisions, traditionally associated with damping, to enhance flavor coherence is now explained: it is a temporary growth that serves to reduce the system's free energy on its way toward a thermodynamically more favorable state.

\section{Discussion and summary}\label{sec:discussion}

Our main goal was to develop a comprehensive understanding of the dispersion relation of the neutrino plasma and to elucidate the connections between fast, slow, and collisional instabilities. Such connections are expected due to a distinct hierarchy in the collective flavor conversion problem: vacuum frequencies and collision rates are much smaller than the refractive energy shift. Therefore, they must act as perturbations, and the instabilities they cause can be physically understood within this framework.

To summarize our findings, fast instabilities uniquely arise from an angular crossing; they originate from Landau-damped modes which, within a narrow range -- corresponding to flavomons resonant with neutrinos in the ``flipped'' region -- can become anti-damped, and thus unstable. In contrast, slow and collisional instabilities do not originate from Landau-damped modes. Slow unstable modes arise from flavomons moving nearly at the speed of light; as such, they resonate only with collinearly moving neutrinos and, by virtue of the energy splitting between neutrinos and antineutrinos, can develop an effective growth rate. We have also generalized our previous findings, based on monochromatic energy distributions, to arbitrary energy distributions.

Collisional instabilities do not arise from a wave-particle resonance, but rather from a direct attempt by the system to lower its free energy. An alternative viewpoint is that neutrino collisions enable flavomon production by lifting the exact requirement of energy conservation, analogous to bremsstrahlung emission of photons in a plasma when the Cherenkov condition is not met and electrons would not radiate in the collisionless limit. We have found that if the dominant species among $\nu_e$ and $\overline{\nu}_e$ collides less frequently than the other, collisional instability arises for gapped modes, defined by $\mathrm{Re}(\omega)\sim \mu \epsilon$. This case might be relevant for neutron star mergers, where typically there are more $\overline{\nu}_e$ than $\nu_e$, at least in some regions, while the interaction rate for $\nu_e$ is generally larger than for $\overline{\nu}_e$. On the other hand, in SNe we usually have the opposite situation, with a dominance of $\nu_e$ over $\overline{\nu}_e$. In this case, collisional instability arises for gapless modes, which have $\omega\sim \Gamma_E$. However, if the neutrino-antineutrino asymmetry becomes sufficiently large, these unstable gapless modes may disappear altogether; we have identified the exact condition for this to happen for a generic distribution.

For the first time, we have introduced clearly the concept of gapped and gapless modes and we now comment on the physical implications of this distinction. Gapped modes are defined as having a large eigenfrequency $\mathrm{Re}(\omega)\sim \mu \epsilon$, and represent weak perturbations of the corresponding fast eigenmodes  in the limit of vanishing $\tomegaE$ and $\Gamma_E$.  Therefore, $\omega$ changes considerably over intervals of wavenumber $\delta k\sim \mu \epsilon$. In contrast, gapless modes arise from a dispersion relation that, to first approximation, does not contain $\mu$ at all (see~Eq.~\ref{eq:dispersion_relation_slow_only}), and thus their frequency varies considerably already over $\delta k \sim \tomegaE$ or $\Gamma_E$. This makes a crucial difference in inhomogeneous media.

The dispersion relation approach assumes perfect homogeneity, whereas realistic astrophysical environments display large-scale gradients over distances $\ell$, which can be on the order of hundreds of meters. (Here, we do not refer to small-amplitude turbulent fluctuations of the matter density.) Such inhomogeneities mix modes with wavenumbers differing by $\Delta k \sim \ell^{-1}$. For gapped modes, which originate from fast modes with very short wavelengths, this mixing is likely a small perturbation, whereas for gapless modes, it can represent a large effect. Therefore, in this context, the dispersion relation may be an unreliable method for studying gapless modes; a realistic large-scale inhomogeneous setup would likely be needed -- a question that goes beyond the tools developed here. This is especially important since collisional modes are primarily gapless in SN environments, although they may transition to near-luminal gapped modes and still remain unstable.

Our main takeaway is that slow modes are unstable only when they are near-luminal and gapped, whereas collisional modes may become unstable when they are gapless, specifically in the physical regime where neutrinos scatter more rapidly than antineutrinos. These gapless modes may transition into unstable near-luminal gapped modes; however, collisional instabilities of this kind appear only when the asymmetry between neutrino and antineutrino densities is sufficiently small. If neutrinos and antineutrinos collide with very different rates, even asymmetries of order 1 can produce the collisionally unstable gapless branch, which at larger $k$ turns into a near-luminal unstable branch. The existence of modes with very small $\mathrm{Re}(\omega)$ at $k=0$ was already recognized in previous studies of collisional instabilities. We have shown here that their properties are much more general than previously suggested, and that they belong to the general branch of modes existing only at very small $k$. However, as emphasized earlier, gapless modes are the least robust prediction of the dispersion relation in the presence of large-scale gradients. In contrast, the possibility of near-luminal gapped modes was identified here for the first time. 

The near-luminal nature of unstable modes -- both of the slow and potentially of the collisional type -- has important physical consequences. As discussed earlier~\cite{Fiorillo:2025ank}, unstable flavomons move with nearly the speed of light, suggesting that the instability is convective, as explicitly proven for slow instabilities. As a result, the produced flavomons leave their region of origin, making collective flavor evolution intrinsically non-local. Going one step further, for any slow or collisional unstable mode, the growth occurs over large length and time scales, even though the typical wavelengths themselves are short. Slow and collisional instabilities are therefore intrinsically multi-scale problems, and the decoupling of scales -- often advocated for fast modes in works attempting to describe fast evolution in small volumes, e.g. Refs.~\cite{Bhattacharyya:2020jpj, Zaizen:2022cik, Nagakura:2022kic, Xiong:2023vcm, Shalgar:2022lvv, Cornelius:2023eop, Abbar:2024ynh, Richers:2024zit} -- presumably breaks down completely. In this case, the relaxation of instabilities cannot be studied in a small volume and then extrapolated to large scales using numerically informed recipes. Even for fast modes, we have questioned the idea of local relaxation~\cite{Fiorillo:2025ank}, since for weak instabilities the produced flavomons move nearly at the speed of light toward the direction of the “flipped” region of the angular distribution, and thus can affect regions far away from their production. 

Our study was limited to longitudinally polarized modes, leaving out the axial-breaking ones, which, however, do not require a dedicated investigation. In the slow case, we have already shown~\cite{Fiorillo:2024pns} that slow instabilities can arise once again near the light cone, where the Landau-damped modes merge with the stable superluminal branch. In the collisional limit, it is instead the superluminal modes that become unstable, exactly as for the longitudinal ones; it is straightforward to repeat the analysis of Secs.~\ref{sec:superluminal} and~\ref{sec:gapless} for axial-breaking modes, and no new features arise. Another simplification was assuming axial symmetry for the overall angular distribution. However, our main result is that the phenomenologically most interesting unstable modes are near-luminal in both the slow and collisional limits. For these modes, the instability depends primarily on the angular distribution evaluated in the direction of the mode, i.e., collinear with the flavomon. Therefore, our results can be immediately extended to general angular distributions by simply replacing $G(1,E)$ and $\oG(1,E)$ with the corresponding distributions evaluated along this direction.

While our derivations may sometimes seem rather theoretical, this work is actually driven by practical questions that can be answered. Without solving the dispersion relation, one can comprehensively determine the order of magnitude of the growth rate for different regimes of instability: slow or collisional instabilities have growth rates of the order of $\tomegaE/\epsilon$ or $\Gamma_E/\epsilon$ respectively  in the broad-resonance regime, i.e., when $\epsilon \ll 1$, unless $\tomegaE\gtrsim \mu\epsilon^2$ or $\Gamma_E\gtrsim \mu\epsilon^2$, in which case the growth rate is of order of $\sqrt{\tomegaE\mu}$ or $\sqrt{\Gamma_E\mu}$. The first regime is very likely the physically relevant one. In this regime, collisional instabilities extend from $k\sim 0$ to $k\sim \mu \epsilon$, so unstable small-scale modes can never be neglected. Slow instabilities exist only for $k\sim \mu\epsilon$, since $k\sim 0$ is stable in this regime. In contrast, fast instabilities have growth rates of order $\mu\epsilon$, but these are generally proportional to the number of neutrinos in the ``flipped'' region of the angular crossing~\cite{Fiorillo:2024bzm,Fiorillo:2024uki}, so that for weak angular crossings it can be considerably smaller. 

Similarly, when $\epsilon \simeq 1$, corresponding to a weak \textit{energy} crossing in which the DLN is similar to the total amount of neutrinos in the system, the slow instability can also be weak, which we have denoted here as narrow-resonance because it leads to rather small growth rate and correspondingly rather well-defined flavomon energy levels. The typical values of the growth rate in this case can be inferred from Eq.~\eqref{eq:approximate_growth_rate_slow_resonant}, which is completely analytical and we have shown to be extremely accurate. As for collisional instabilities of the gapless kind, which are the most interesting for SNe, where we have a dominance of $\nu_e$ over the other flavors, in the limit $\epsilon \simeq 1$ they seem to disappear altogether, and we have determined the threshold condition for their disappearance.

This paper concludes our series of studies on the general dispersion theory of the neutrino plasma~\cite{Fiorillo:2024bzm, Fiorillo:2024uki, Fiorillo:2024pns, Fiorillo:2025ank}. We have shown that fast and slow instabilities are fundamentally of the same nature, arising from a general wave-particle resonance picture. We have also developed a consistent framework to estimate the approximate orders of magnitude for the growth rates of instabilities across all classes, depending on their regime of appearance. These findings provide an intuitive understanding of the phenomenology of the instabilities and clarify which regions of realistic astrophysical systems may be significantly affected. Our general treatment of the linear regime of flavor instabilities lays the foundation for studying subsequent non-linear evolution, which may be most effectively addressed within the paradigm of neutrinos interacting with flavomons, thereby circumventing otherwise unavoidable small-scale challenges.

\section*{Acknowledgments}

We acknowledge useful conversations with Julien Froustey, Luke Johns, and Ian Padilla-Gay.
DFGF is supported by the Alexander von Humboldt Foundation (Germany). GGR acknowledges partial support by the German Research Foundation (DFG) through the Collaborative Research Centre ``Neutrinos and Dark Matter in Astro- and Particle Physics (NDM),'' Grant SFB-1258-283604770, and under Germany’s Excellence Strategy through the Cluster of Excellence ORIGINS EXC-2094-390783311.

\bibliographystyle{JHEP}
\bibliography{References}

\newpage

\section*{Post-publication Erratum}
\addcontentsline{toc}{section}{Post-publication Erratum}

\subsection*{Typographical Error in Eqs.~(2.12) and (2.14)}

We have become aware of a typographical error in Eqs.~\eqref{eq:2.12} and 
\eqref{eq:2.14}, where $+\lambda$ incorrectly appeared in the denominators, even though it was already absorbed in the shifted frequency $\omega$. (Thanks to Amitava Dasgupta and Sovan Chakraborty.)

\subsection*{Deformation of Phase-Space Integration Contour}

The dispersion relation of flavor waves depends on neutrino phase-space integrals that involve a potentially resonant denominator $(\omega-k v-w)^{-1}$, where $\omega$ is the flavor-wave frequency and $k$ its wave number along the symmetry axis. The integration is over neutrino energy encoded in $w$ which we here use for $(\delta m_\nu^2/2E)\sin(2\theta_{\rm V})$ and over $v=\cos\theta$, the neutrino direction along the symmetry axis. For unstable modes, defined by $\omega$ having a positive imaginary part, the integrals $\int_{-\infty}^{+\infty}dw\int_{-1}^{+1}dv$ along the real axis of $w$ and $v$ can be performed without problem, yielding the unstable normal modes. 

For damped modes, on the other hand, the result depends on the deformation of the integration paths in the complex planes of $w$ and $v$. In our presentation, and in deriving some of the damped modes numerically, we have performed the correct deformation for the integral over $dv$, but then integrated over $dw$ along the real axis without deformation. This may lead to an integration through a branch cut and can cause one to miss some of the Landau-damped modes. 

In a recent paper~\cite{Kost:2026jrk}, Kost and Duan have circumvented this problem by transforming the phase-space integral to new variables such that only in one of them a deformation is required. In their Fig.~6, they show explicit examples for the additional Landau-damped modes that appear with the correct deformation procedure. Of course, all modes also appear when using the original variables combined with the appropriate deformation of both integration paths. 

The main conclusions of our paper remain unaffected, whereas additional Landau-damped modes might appear in some examples. 

In the following, we explain more explicitly these very technical but important issues. We consider a general structure for the integrals
\begin{equation}
    I(\omega,k)=\int_{-\infty}^{+\infty}dw \int_{-1}^{+1}dv \frac{g(w,v)}{\omega-kv-w}
\end{equation}
appearing in the dispersion relation. 
One way to do these integrals, adopted by Kost and Duan~\cite{Kost:2026jrk}, is to change the integration variable to $w'=w+kv$
\begin{equation}
    I(\omega,k)=\int_{-\infty}^{+\infty}\frac{dw'}{\omega-w'}\int_{-1}^{+1}dv\,g(w'-kv,v).
\end{equation}
The integral over $v$ is then trivial, while the integral over $w'$ is performed over a complex path that always passes below the singularity $w'=\omega$, following the Landau prescription~\cite{Landau:1946jc} recently exported to collective flavor evolution~\cite{Fiorillo:2023mze,Fiorillo:2024bzm}.

In our paper, we instead proceeded differently, and performed first the integration over~$v$ 
\begin{equation}
    F(w;\omega,k)=\int_{-1}^{+1}dv \frac{g(w,v)}{\omega-kv-w},
\end{equation}
which we correctly continued analytically by deforming the contour in $v$ to pass below the singularity $v=(\omega-w)/k$ (assuming $k>0$). We then defined the integral as
\begin{equation}
    I(\omega,k)=\int_{-\infty}^{+\infty}dw F(w;\omega,k).
\end{equation}
However, as noted by Kost and Duan~\cite{Kost:2026jrk}, the function $F(w; \omega,k)$ is not entirely analytical, but possesses two branch cuts in the complex plane of $w$ originating from the two branch points $w_\pm=\omega\pm k$. These branch points originate from the finite integration range in $v$; their branch cuts can be chosen to point upwards with $\mathrm{Im}(w)>\mathrm{Im}(\omega)$ and $\mathrm{Re}(w)=\mathrm{Re}(w_\pm)$.

\begin{figure}
    \includegraphics[width=\textwidth]{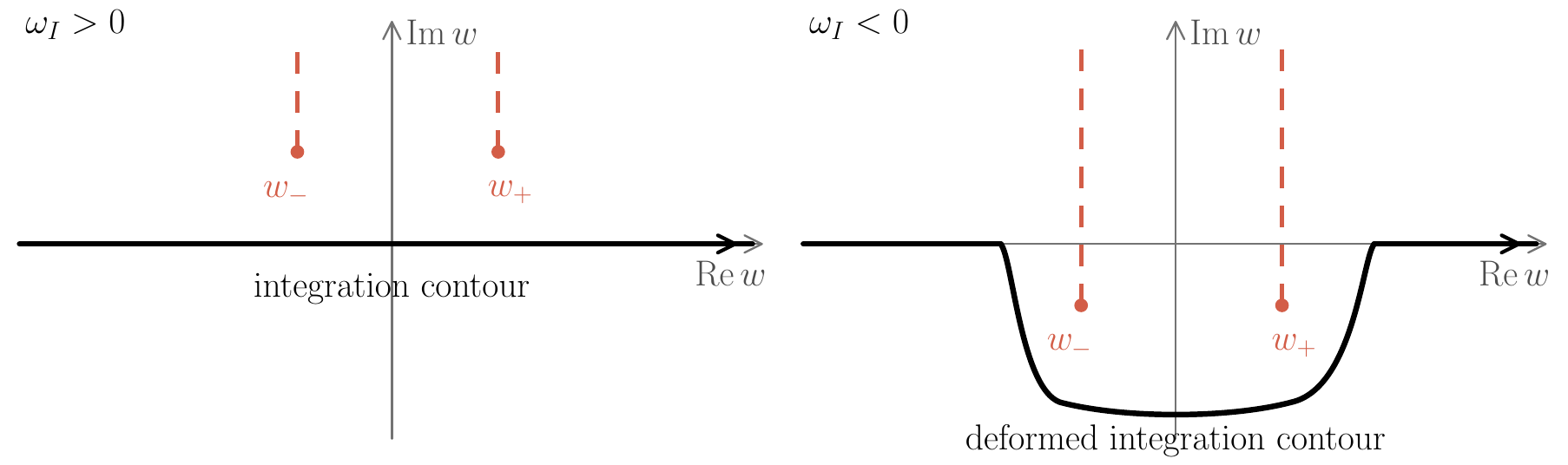}
    \caption{Singularity structure of $F(w;\omega,k)$ in the complex plane of $w$. The two points $w_\pm$ defined in the text are branch points, due to the integration over a finite range of $v$. For $\omega_I=\mathrm{Im}(\omega)>0$, this causes no issue; however, for $\omega_I<0$, the contour of integration over $w$ should be deformed so as to pass below the branch points.}\label{fig:deformation}
\end{figure}

The integration over $w$ can still be performed, by deforming the contour of integration in $w$ so as to pass below the branch points $w_\pm$. The deformation is schematically shown in Fig.~\ref{fig:deformation}. When this procedure is correctly applied, the integration with this second method leads to the same result as the method of Kost and Duan, although it is perhaps more cumbersome in its practical application. Generally, since the integral is defined as an analytical continuation, the order in which the integrals are performed makes no difference in the final result.

In our paper, we did not explain this difference, and in fact the results obtained in our figures were obtained \textit{without} applying this deformation. The consequence is that some Landau-damped solutions with $\mathrm{Im}(\omega)<0$ may not have been correctly identified. Kost and Duan report an explicit example of this failure. 

On the other hand, this shortcoming does not affect the main results of our work and in particular its message, namely that it is possible to determine the unstable bands, those with $\mathrm{Im}(\omega)>0$, by analytical expansions of the retarded dielectric function for small $|\mathrm{Im}(\omega)|$. The presence of strongly Landau-damped modes is not a direct concern of our work. It is actually rather difficult, or perhaps impossible for a general distribution, to find \textit{all} of the Landau-damped modes: A branch of purely Landau-damped modes may be entirely disconnected from the unstable band (which can be determined analytically from our methods), in which case it might only be found by trial-and-error. Kost and Duan manage to determine all of the solutions for a special family of Gaussian distribution functions, for which the integrals can be performed analytically and the solution is exact. 

We should also stress that the presence of strongly Landau-damped modes generally need not be connected with physical behavior, since their nature depends on the behavior of the distribution function $g(w,v)$ at complex values of $w$ and $v$. While mildly Landau-damped modes depend only on this continuation immediately below the real axis, strongly Landau-damped modes rely on the continuation significantly below the real axis. Two functions which look very similar on the real axis may look entirely different far from it. This is of course in agreement with the physical observation that strongly Landau-damped modes are generally unobservable in practice, since the evolution will be dominated by stable, unstable, or mildly Landau-damped modes, or by the long-term response connected with the branch cuts in $\omega$. For this reason, we do not attempt a complete characterization of the Landau-damped modes, and focus only on the modes close to the unstable ones.

\end{document}